\documentclass[
aps, 
amsmath,
amssymb,
reprint,
superscriptaddress
]{revtex4-2}

\usepackage{graphicx}
\usepackage{dcolumn}
\usepackage{bm}
\usepackage{booktabs,tabularx,graphicx,siunitx,float,physics,subcaption}
\usepackage{enumitem}
\usepackage[font=footnotesize,labelfont=bf]{caption}
\usepackage[colorlinks=true,citecolor=blue,urlcolor=blue]{hyperref}

\begin{document}


\newcommand{\papertitle}{Evidencing Macroscopic Quantum Coherence using Confined Uniform Magnetic Field}

\title{\papertitle}

\author{Dipankar Home}
    \thanks{These authors contributed equally to this work.}
    \affiliation{Raman Research Institute (RRI), C.~V.~Raman Avenue, Sadashivanagar, Bengaluru, Karnataka 560080, India}
    \affiliation{Bose Institute, Kolkata, West Bengal 700091, India}

\author{Harsh Talwar}
    \thanks{These authors contributed equally to this work.}
    \affiliation{Indian Institute of Science Education 
    \& Research (IISER), Kolkata, Mohanpur, West Bengal 741246, India}

\author{Nilaj Saha}
    \affiliation{Indian Institute of Science Education \& Research (IISER), Mohali, SAS Nagar, Punjab 140306, India}

\author{Sougato Bose}
    \affiliation{University College London, Gower Street, London WC1E 6BT, United Kingdom}

\author{Arka Banerjee}
    \affiliation{Indian Institute of Science Education \& Research (IISER), Pune, Maharashtra 411008, India}


\begin{abstract}

A previously unexplored scheme is devised for creating and certifying coherent quantum superposition of distinctly separated states of a spin-tagged massive object passing through a confined uniform magnetic field. This results in entangling the spatial and spin parts of a two-component Pauli wavefunction such that the entailed coherence can be conveniently certified by spin measurements alone. The implementability of this scheme is analysed using NV-centre nanodiamonds of masses $\sim 10^{10}$ amu based on the current technological capabilities.

\end{abstract}

\maketitle



\textit{Introduction}--- Among the major unresolved fundamental questions in quantum mechanics is the one concerning the specification of the quantum--classical boundary 
\cite{Sch1935,einstein1949remarks,BohmQuantumTheory,BohmHileyUndivided,Zurek1991,HomeConceptualFoundations,HomeWhitaker,Schlosshauer}. In this context, it is important to devise empirically implementable means for testing the hallmark of quantum mechanics at the macroscopic level, viz.~the quantum coherence embodied in the macroscopic superposition of states \cite{Leggett1980,LeggettSchrodingerCat,Leggett1983Superposition,Leggett1986CurrentStatus}. To this end, the Leggett–Garg inequality (LGI) \cite{LGineq,leggett2008_lgi} and its variants \cite{kofler2013_lgi,Saha2015WignerLGI} have stimulated a range of studies \cite{Leggett2002TestingLimits,knee2012_lgi,Gangopadhyay2013KaonsLGI,robens2015_lgi,knee2016_lgi,formaggio2016_lgi,Mal2016MacrorealismLargeSpin,Halliwell2022LGI,joarder2022_lgi,debarshi2024_lgi} aimed at probing macroscopic quantum coherence based on the incompatibility between quantum mechanics and macrorealism, evidenced through the quantum mechanically predicted violations of LGI. 
Only recently, these ideas have been extended to arbitrarily massive oscillators \cite{debarshi2024_lgi}. However, experimental demonstrations of macroscopic quantumness for the spatial superposition of states of large masses are still lacking; matter-wave interferometric tests of quantum superposition principle for spatially separated states have so far been realized only for masses up to $10^4$--$10^5$ amu \cite{FEIN2019,Pedalino2026NanoparticleInterferometry}.
While quantum superposition has been demonstrated for a $16~\mu\mathrm{g}$ mechanical resonator \cite{Bild2023SchrodingerCat}, this involved superposed states entailing two opposite-phase oscillations, rather than a superposition of spatially separated localized states. Hence, fresh ideas feasible for near-term experimental implementation are called for to facilitate tests of spatial superposition of states for \textit{increasingly large} masses.

To this end, in the present paper, we initiate a distinct direction of study for demonstrating Macroscopic Quantum Coherence (MQC) using a \textit{massive} system. In particular, we focus on testing the superposition of \textit{spatially separated} localized wavepackets whose separation between their respective peaks is greater than the respective widths of the superposed individual wavepackets. Such spatial superpositions of distinctly localised states of large masses have both fundamental and practical implications, ranging from providing empirical constraints on the models of wavefunction collapse \cite{Bassi2003,romero2011,bassi2013}, evidencing quantum feature of gravity in tabletop experiments \cite{sougato_2017,marletto_2017,belenchia,sougato_2024,sougato25_rev_mass}, to sensing extremely weak classical gravity~\cite{Marshman2020MIMAC}. A commonly pursued approach for this purpose relies on Stern-Gerlach (SG) interferometry. For atomic systems, spatial coherence has been demonstrated via SG momentum splitting and observation of spatial interference fringes after postselecting a specific spin state~\cite{Machluf2013,Margalit2019SGInterferometer}, but such approaches have not yet been scaled to massive systems. A proposed but yet unrealized alternative route is to certify spatial coherence through the entanglement between the spin and the center-of-mass motion of an object in a SG setup~\cite{massive_qubits_Bin23}. A third approach relies on full-loop SG interferometry~\cite{marshman2022}, where spin-dependent splitting by an inhomogeneous magnetic field is followed by recombination of the wavepackets to achieve coherent overlap and complete the interferometric sequence, as in a recent experiment using $^{87}$Rb atoms~\cite{SGQGtest1}. However, coherent recombination in such setups requires stringent control over the magnetic-field inhomogeneity~\cite{englert1988,Schwinger1988,Scully1989,Englert1997_time_reversal}. Moreover for realizing the SG setup in itself, the constraints imposed by Maxwell’s equations~\cite{Scully1987_SGA}, together with noise and dissipative effects~\cite{SG_dissipation_06}, make producing and maintaining the required precisely controllable magnetic field gradients a formidable experimental challenge~\cite{SG_reduction_15}.

In contrast, the setup that we shall explore in this paper is based on a spatially confined \textit{uniform} magnetic field, with a generic spin-1/2 system passing through it. Note that spin is fundamentally described by the relativistic Dirac equation through the spinor representation of the Lorentz group. In the regime where the kinetic energy ($K$) and Zeeman energy ($|\mu B|$, $\mu$ being the magnetic moment of the system) are small compared to the rest-mass energy ($mc^2$), the Dirac equation systematically reduces to the two-component Schrödinger-Pauli equation~\cite{LancasterQFT}. Our analysis is focused on this non-relativistic regime ($|\mu B| < K \ll mc^2$), where the Pauli equation appropriately captures the coupled spatial--spin dynamics. Within this non-relativistic regime, in the further limit $|\mu B| \ll K$, the dynamics in our setup reduces to the accumulation of a relative phase between the spin components, equivalent to the standard Larmor precession result~\cite{home2013}. Notably, unlike the textbook treatment of Larmor precession where the spin-1/2 particle is typically considered to be at rest, thereby ignoring its spatial evolution altogether~\cite{qm_textbooks}, our framework fully incorporates the spatial dynamics and captures the resulting nontrivial spatial--spin coupling effects.


A setup closely related to ours has previously been investigated in the important context of computing and measuring the so-called tunneling times through a potential barrier~\cite{PhysRevB.27.6178,steiberg_rev,tunnelling_2017_jul,tunnelling_2017_dec,Ramos2020TunnellingTime,tunnelling_2026}. However, in this present paper, we focus on the unexplored possibility of harnessing the spatial--spin coupling in this setup to generate and certify MQC in the spatial degree of freedom. In particular, we uncover how the spatial-spin entangled form of the emergent wavefunction manifests in the measurable spin statistics, and identify the experimentally accessible parameter regimes in which the genuine quantum nature of the \textit{spatial motion} can be probed using \textit{only} spin measurements. One of the key feature of our approach is its robustness to realistic smooth-edged magnetic field profiles while probing quantum spatial coherence for progressively larger masses. \vspace{2pt}


\textit{The basics of our treatment}--- We begin by considering a \textsl{generic} electrically neutral spin-$\tfrac{1}{2}$ system of mass $m$, initially spin-polarised along $+x$, moving along the $+\hat{x}$-axis through a finite region of width $w$, containing a uniform magnetic field, $B(x)$ oriented along $+\hat{z}$ (see Fig.~\ref{fig:setup}). The dynamics is obtained by evolving the initial two-component spinor using the appropriate Pauli equations which decompose into two uncoupled equations for the spatial wavefunctions $\psi^{\pm}$ corresponding to the $\ket{\uparrow}_z$ and $\ket{\downarrow}_z$ spin components respectively,
\begin{equation} \label{eq:Pauli_eqs & Hamiltonian}
   H_{\pm} \, \psi^{\pm} = i \hbar \frac{\partial \psi^{\pm}}{\partial t} \,; \quad H_{\pm} = -\frac{\hbar^2}{2m} \frac{\partial^2}{\partial x^2} \mp \mu B(x) \,.
\end{equation}

\noindent The initial center-of-mass state is modelled to be a Gaussian wavepacket of width $\sigma$, centered at $x_{0}$, such that
\begin{equation}\label{eq:tot_ini_state}
    \begin{gathered}
    \Psi_{i}(x,0) = \psi_{0}(x,0) \otimes \frac{1}{\sqrt{2}}(\ket{\uparrow}_z + \ket{\downarrow}_z) \,, \\
    \text{where} \,\, \psi_{0}(x,0) = \frac{1}{(2\pi\sigma^2)^{1/4}} e^{i k_0 x} e^{-\frac{(x-x_0)^2}{4\sigma^2}} \,.
    \end{gathered}
\end{equation}
Defining $r \equiv |\mu B_{0}|/K$, with $B_{0}$ denoting the uniform magnetic-field strength and $K$ the incident kinetic energy, we obtain $k_0 = \sqrt{2m |\mu B_{0}| / r \hbar^{2}}$. Then, in accordance with Eq.~\ref{eq:Pauli_eqs & Hamiltonian}, the time-evolved complete wavefunction, at $t=t_{f}$, emerging after interacting with the confined magnetic field is of the form,
\begin{equation} \label{eq:state_at_tf}
    \begin{aligned}
    &\Psi_{q}(\mathbf{x}, t_{f}) 
    = \exp\!\left(-\tfrac{i}{\hbar} H t_{f}\right) \Psi_i (\mathbf{x},0)\\
    &\, = \frac{1}{\sqrt{2}} \Big[ \psi_T ^{+}\!(\mathbf{x},t_{f}) \otimes \ket{\uparrow}_{z}
    + \psi_T ^{-}\!(\mathbf{x},t_{f}) \otimes \ket{\downarrow}_{z} \Big] \,.
    \end{aligned}
\end{equation}
Here, Eq.~\ref{eq:state_at_tf} exhibits intraparticle entanglement~\cite{intraparticle_review_2020} between the spin and spatial degrees of freedom. This arises from the combined action of the kinetic energy term and the spatially confined spin--field interaction. We call this final state as the \textit{spatial-spin coupled} wavefunction.
\begin{center}
\vspace{1mm}
\includegraphics[width=0.48\textwidth]{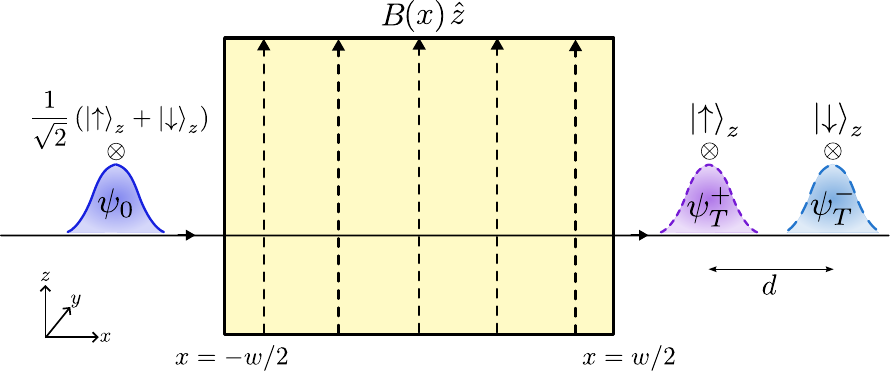}
\captionof{figure}{Schematic of the arrangement. $\psi_{T}^{\pm}$  are the transmitted spatial components corresponding to the spin components $\ket{\uparrow}_{z}$ and $\ket{\downarrow}_{z}$ respectively. $B(x)$ is a uniform magnetic field confined within $x \in \{-w/2,w/2\}$ and directed along $+\hat{z}$.}
\label{fig:setup}
\end{center}
Another key feature embodying the crux of our analysis is the following: From Eq.~\ref{eq:Pauli_eqs & Hamiltonian}, it is seen that one of the spatial components (depending on the sign of $\mu$) of the spinor essentially evolves under a \textit{potential barrier}, while the other component evolves under a \textit{potential well} due to the nature of the spin-magnetic field interaction. Consequently, the two components acquire different velocities within the magnetic field region, resulting in a spatial separation \textit{along the direction of propagation} (unlike SG-interferometry setups). This separation remains fixed with time after the system leaves the magnetic field region and undergoes free evolution, resulting in the creation of a superposition of spatially separated localised states whose coherence is testable in terms of the predicted spin statistics. \vspace{2pt} 

\textit{Analysis}--- We begin by defining the main parameters that are crucial ingredients for the certification of MQC in terms of the spatial degree of freedom. To this end, we note that quantum spatial coherence is naturally denoted by the squared inner product between the outgoing spatial components, $ \abs{\braket{\psi^{+}_{T}}{\psi^{-}_{T}}}^2$. Although this quantity is specified in terms of the spatial part of the wavefunction, the spatial-spin entanglement entailed by our treatment allows it to be expressed entirely in terms of the Pauli spin observables $\sigma_x ~\text{and}~ \sigma_y$ (see~\ref{app:analysis_coh_param} for details), so that we can define the \textit{coherence parameter} `$\mathcal{O}$' as
\begin{equation} \label{eq:O_main}
    \mathcal{O} \equiv \expval{\sigma_x}^2 + \expval{\sigma_y}^2 .
\end{equation}
Here, $\mathcal{O}$ directly encapsulates the off-diagonal elements in the reduced spin density matrix of the final wavefunction and is a measure of the overlap between the separated spatial components. Another parameter of key interest is the spatial separation `$d$' between the Gaussian peaks of the two transmitted spatial components, $\psi^{\pm}_{T}$ of $\Psi_{q}$. This quantifies the size of the resulting superposition of macroscopic states, in the sense that we can characterize the extent of spatial delocalisation through a dimensionless \textit{delocalisation parameter}, $d/\sigma$.

\noindent Hereafter, we divide the analysis into three parts: (i) quantum treatment for a sharp potential, (ii) quantum treatment for a smooth potential, and (iii) semiclassical treatment. Throughout our treatment, wavepacket dispersion is taken to be negligible over the relevant evolution times.\vspace{2pt}

(i) \textit{Sharp potential}: The sharp-edged field is modelled as $B(x) = B_0$ for $|x| < w/2$, and $B(x) = 0$ elsewhere. Then, using the total transmitted spatial--spin coupled wavefunction, $\ket{\Psi_{q}} = \frac{1}{\sqrt{2}} \left( \ket{\psi^{+}_{T}} \otimes \ket{\uparrow}_z + \ket{\psi^{-}_{T}} \otimes \ket{\downarrow}_z \right)$ (see~\ref{app:analytical} for detailed expressions of the transmitted wavepacket components $\ket{\psi^{\pm}_{T}}$), we calculate our coherence parameter to be
\begin{equation} \label{eq:coh_para_O_Mainbody}
\mathcal{O} \approx |C_{+}|^{2} |C_{-}|^{2} \, e^{-d^{2}/4\sigma^{2}} \,.
\end{equation}
Here, $|C_{\pm}|^{2}$ are determined by the transmittance of the wavepacket through the field, and incorporate the effect of wavepacket reflection. We further derive the spatial separation $d$ as a function of $r$ (see~\ref{app:gaussian_coh_para,etc}),
\begin{equation} \label{eq:RS3_dist}
d = w \left( \frac{1}{\sqrt{1-r}} - \frac{1}{\sqrt{1+r}} \right) \,.
\end{equation}
Here, note that the parameter $d$ is a strictly increasing function of $r$.
Next, to assess the practical implementability of our proposal, it is crucial to verify to what extent the predictions obtained for the sharp potential remain robust for realistic smooth-edged magnetic field profiles. \vspace{2pt}

(ii) \textit{Smooth potential}: The smooth-edged field is modelled as $B(x) = \frac{B_0}{2}\left[\tanh\!\left(s\left(\frac{w}{2}+x\right)\right) + \tanh\!\left(s\left(\frac{w}{2}-x\right)\right) \right]$. Here, the parameter $s$ controls the smoothness of the field edges, such that smaller values of $s$ correspond to smoother-edged fields. The final transmitted wavefunction is obtained using the WKB approximation (valid when the wavenumbers of the spatial components satisfy $|dk_{\pm}/dx| \ll k_{\pm}^2$) and by assuming dominant transmission (for a qualitative justification see \ref{app:probs_analysis_smooth}). Both the dominant transmission assumption and WKB approximation are well satisfied for the parameter regimes considered here. This approach retains spatial-spin coupling effects and captures the principal effect of the separation of the two spatial components (for details, see~\ref{app:approx}).

\noindent Then, the spin-dependent scattering phases induced by a smooth position-dependent magnetic field can be described within this approximation as
\begin{equation}
\delta_{\pm}(k) = \int \!\left[\sqrt{k^2 - \tfrac{2m}{\hbar^2} V_{\pm}(x)} - k \right] dx \,\,,
\end{equation}
leading to a relative phase $\phi = \delta_{+} - \delta_{-}$ between the spinor components. For a Gaussian incident packet, the final transmitted spatial-spin coupled wavefunction is
\begin{equation}\label{eq:wkb_soln_main}
\begin{split}
    \ket{\Psi_{q}(x,t)} &= \frac{1}{\sqrt{2}}\left(\psi_T ^{+}(x,t) \, \ket{\uparrow}_{z} + \psi_T ^{-}(x,t) \, \ket{\downarrow}_{z}\right) \\ 
    \psi_T^{\pm}(x,t) &= \frac{1}{(2\pi\sigma^2)^{1/4}} e^{i(k_0x-\omega_0 t+ \delta_\pm(k_0))} e^{-\frac{(x - x_{\pm}(t))^2}{4\sigma^2}}
\end{split}
\end{equation}
where \( x_{\pm}(t) = x_0 + v_g t - (d\delta_{\pm}/dk)|_{k=k_0} \) (with \( v_g = \hbar k_0/m \)). The transmitted spinor separates into two subpackets with a relative phase difference of $\phi_0 = \delta_-(k_0) - \delta_+(k_0) $. and a separation of $d=\left (d\delta_+/dk)\right|_{k=k_0}-\left.(d\delta_-/dk)\right|_{k=k_0}$. This expression for $d$ reduces to Eq.~\ref{eq:RS3_dist} in the large $s$ limit (i.e.~for a sharp-edged field). Furthermore, the inner product comes out as $\braket{\psi_{T}^{+}}{\psi_{T}^{-}} = e^{i\phi_0}\,e^{-d^2/8\sigma^2}$, resulting in $\mathcal{O} = e^{-d^2/4\sigma^2}$. \vspace{2pt}

(iii) \textit{Semiclassical treatment}: A central feature of the preceding analysis is the entanglement between the spin and spatial degrees of freedom of the wavefunction. To establish that the observed spin statistics irreducibly originates from this coupling, it is essential to contrast the coupled treatment with and rule out a scenario in which such coupling is absent; the detailed rationale of this comparison for the purpose of certifying MQC is discussed in~\ref{app:semiclassical}. Accordingly, we consider the regime where the influence of the spin--magnetic field interaction on the spatial dynamics is negligible, so that the spatial mode evolves freely while the spin undergoes conventional Larmor precession under the magnetic interaction. We refer to the corresponding final state in this case as the \textit{semiclassical} wavefunction,
\begin{equation}\label{eq:semiclassical_wavefunc}
    \begin{split}
        \ket{\Psi_{sc}} &= \psi(x,t) \otimes \frac{1}{\sqrt{2}}\left( \, \ket{\uparrow}_{z} + e^{i\phi} \, \ket{\downarrow}_{z}\right) \\ 
        \psi(x,t) &= \frac{1}{(2\pi\sigma^2)^{1/4}} e^{i(k_0x-\omega_0 t)} e^{-\frac{(x - x_0 - v_g t)^2}{4\sigma^2}}
    \end{split}
\end{equation}
where $\phi = \frac{\sqrt{2 |\mu B_{0}| w^{2} m r}}{\hbar}$. From Eq.~\ref{eq:semiclassical_wavefunc}, we note that the coherence parameter remains $\mathcal{O}=1$, independent of the value of $r$. \vspace{1mm}

\textit{Key observations}--- Before proceeding to discuss how coherence is certified, we note the important features which consistently hold for both sharp and smooth potentials. To begin with, see from Eq.~\ref{eq:RS3_dist} that $d$ strictly increases with $r$. Further, $\sigma$ is fixed by the initial wavepacket preparation and is assumed to remain constant. Then, using the delocalisation parameter $d/\sigma$, we can divide the $r$-parameter space into two distinct regimes based on considerations discussed in \ref{app:data1_sigma}: a ``small-$r$" regime defined by $d/\sigma < 5$, and a ``large-$r$" regime defined by $d/\sigma > 5$. Now, the following salient points are worth highlighting: 
\begin{enumerate}[nosep]
    \item For the spatial--spin coupled state, the \textit{coherence parameter} and the \textit{delocalisation parameter} are related in an elegant manner, $\mathcal{O} \approx \exp(-\frac{1}{4}(d/\sigma)^2)$.
    \item In the $d/\sigma \rightarrow 0$ limit of the small-$r$ regime, both the sharp- and smooth-field solutions asymptotically reduce to the semiclassical one, with $\mathcal{O} \to 1$.
    \item As $r$ is increased within the small-$r$ regime, $\mathcal{O}$ undergoes Gaussian decay and the spatial separation increases ($0 < d < 5\sigma$), such that as $d \to 5\sigma$, $\mathcal{O} \to 0$.
    \item Wavepacket reflection remains strongly suppressed throughout the small-$r$ regime.
    \item Next, throughout the large-$r$ regime, $\mathcal{O} \approx 0$. 
    \item We note that these features persist even for increasing system mass, demonstrating the inherent scalability of our scheme to the macroscopic regime.
\end{enumerate}      

\begin{figure}[H]
\centering
\includegraphics[width=\columnwidth]{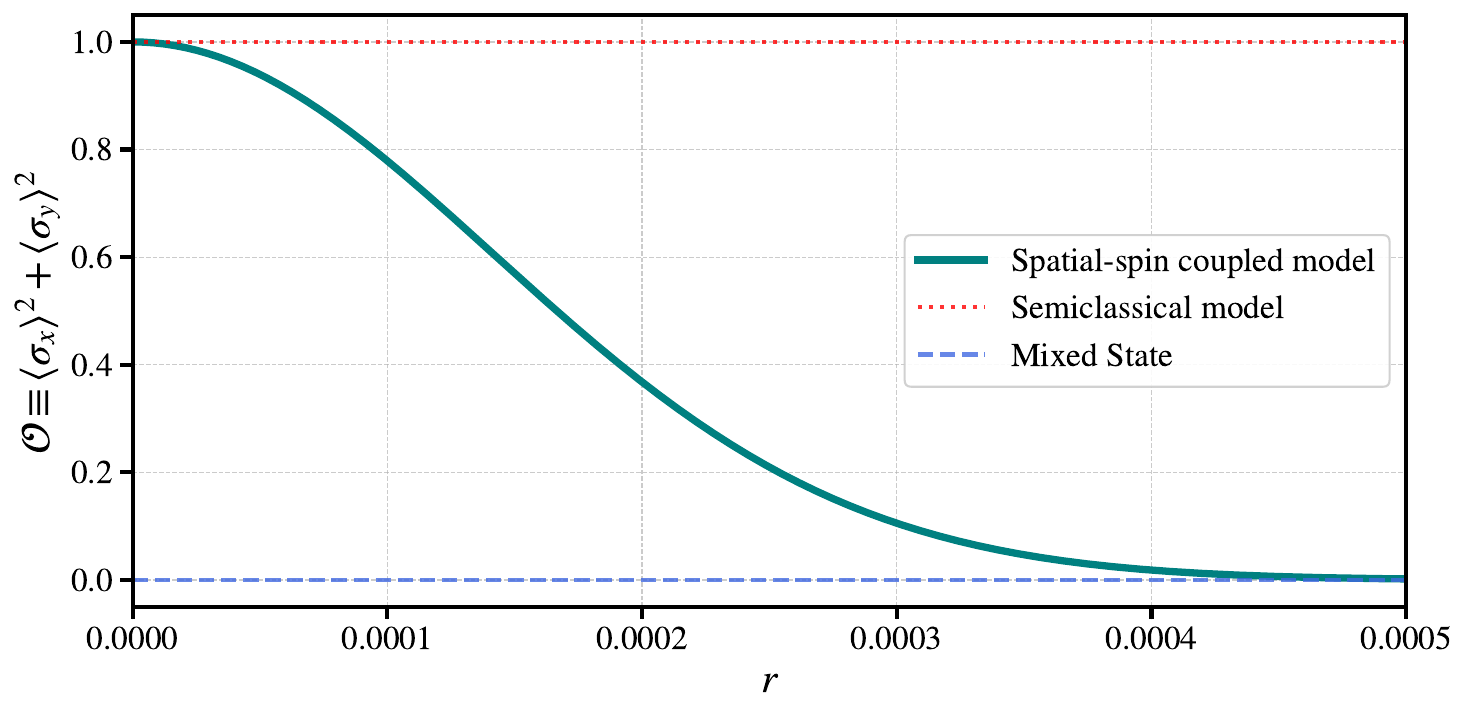}
\vspace{-5mm}
\caption{Coherence Parameter ($\mathcal{O}$) versus $r$ in the small-$r$ regime. The plot holds for both smooth and sharp potentials, demonstrating the robustness of $\mathcal{O}$. Parameters: $m = 10^{7}$ amu, $w =100\,\mu\mathrm{m}$, $s = 0.1\,(\mu\mathrm{m})^{-1}$, $\sigma = 10~\mathrm{nm}$, $B_{0} = 5~ m\mathrm{T}$, $\mu = \mu_{\rm NV}$.}
\label{fig:O_maintext}
\vspace{-3mm}
\end{figure}

\textit{Certifying coherence}--- We propose the criterion: $0 < \mathcal{O} < 1$, to certify MQC in terms of the spatial mode of the transmitted wavefunction. Here, $\mathcal{O}$ is as defined in Eq.~\ref{eq:O_main}, 
making it determinable from spin-only measurements. Crucially, this certification criterion is \textit{not} satisfied by either the semiclassical scenario ($\mathcal{O} = 1$; see~\ref{app:semiclassical}) or incoherent spatial-spin mixed states ($\mathcal{O} = 0$; see~\ref{app:mixed}), which correspond to two \textit{distinct} cases where quantum spatial coherence is absent. Thus, this criterion identifies the relevant range of $r$ values within which genuine spatial coherence is present. Note that $\mathcal{O}$ remains 0 throughout the large-$r$ regime, and hence MQC cannot be certified in that regime using the chosen coherence parameter. Consequently, it is essentially by appropriately tuning the parameter $r$ within the small-$r$ regime, and by confirming our computed trend of $\mathcal{O}$ vs $r$ (see Fig.~\ref{fig:O_maintext}) from the measured spin statistics of $\expval{\sigma_x}$ and $\expval{\sigma_y}$, one would be able to unequivocally certify the inherent quantum spatial coherence arising from the spin-spatial coupling. Furthermore, in the small-$r$ regime, if $w \gg \sigma$, we then obtain $d \approx rw$, independent of the smoothness parameter $s$ even for a smooth field profile. Consequently, $\mathcal{O} ~(\approx e^{-r^2 w^2/ 4 \sigma^2})$ also becomes effectively independent of $s$, demonstrating robustness of both the coherence and delocalisation parameters for the smooth-edged field profiles (see~\ref{app:smooth_limiting}).

Therefore, provided that decoherence effects are adequately suppressed, the measured spin statistics alone can provide a sufficient and robust signature for evidencing MQC. We note, however, that such certification procedure is formulated essentially within quantum theory, and does not rule out alternative non-quantum and non-classical frameworks that may reproduce similar predictions. The key point here is that, for small sized spatial superpositions of distinctly localised states defined by $1< d/\sigma<5$, the certification of coherence within our framework requires neither interferometric recombination~\cite{marshman2022} nor joint position-spin measurements~\cite{massive_qubits_Bin23}, in contrast to conventional SG interferometric tests of large spatial superpositions. \vspace{1mm}

\textit{Implementability of our scheme}--- While our framework applies to a generic massive spin-1/2 system, it can be readily adapted (see \ref{app:analysis_NV}) for the presently much studied system namely the spin-embedded NV center nanodiamonds~\cite{doherty2013}. Before presenting the results for such a system, we first discuss the basic experimental considerations.

\noindent (i) \textit{Magnetic field}: A confined and nearly uniform magnetic field can potentially be engineered over $\mu\mathrm{m}$-scale widths and up to milli-Tesla field strengths
using either micro-Helmholtz coils setups \cite{Kratt2009MicroHelmholtz,Badilita2010MicrocoilsMRI,Spengler2016MicroHelmholtz} or atomic chip based magnetic microtraps \cite{Folman2002AtomChip}.

\noindent (ii) \textit{Coherence time and internal temperature}: The total evolution time $t_f$, including transit through the magnetic field and subsequent free propagation before detection, must be shorter than the intrinsic coherence time ($\tau$) of embedded spins in NV-center nanodiamonds~\cite{nano_2013,nano_2014,NV_coherence_time18,NV_coherence_time22,NV_coherence_time23}. This requires $t_f < \tau$ and $T \lesssim 77~\text{K}$, where decoherence can be well controlled under cryogenic conditions, such that the dominant environmental decoherence channels such as blackbody radiation effects and collision with air molecules~\cite{Fragolino24_decoh,Schut24_decoh,Schut25_decoh} are strongly suppressed~\cite{Romero11_decoh,Pedernales2020}. We emphasize that this constraint involving $t_f$ and $\tau$ imposes a lower bound on the initial velocity ($v_{\mathrm{min}}$) of the system. This consequently limits the maximum admissible mass ($m_{\mathrm{max}}$) for which our scheme can be implemented (see~\ref{app:analytical constraints}). The upshot is that $m_{\mathrm{max}} \propto \mu B_{0} \, \tau^{2}$, implying that improvements in the achievable coherence time or applied magnetic field strength directly enhance the maximum mass that can be prepared in a coherent spatial superposition of states within our framework.

\noindent (iii) \textit{Diamagnetic effect}: Since NV-nanodiamond system is diamagnetic, it is important to note that our treatment of NV-nanodiamond systems does not explicitly incorporate the diamagnetic interaction (spin-independent, but $\textbf{B}^2$-dependent) of the bulk diamond lattice with an external magnetic field. Therefore, our analysis is technically valid in the regime where the Zeeman interaction dominates over the diamagnetic one, constraining the maximum admissible field strength for a given mass; for instance, requiring the diamagnetic term to remain less than one-hundredth of the spin-field interaction term requires $B_{0} \lesssim 5~\mu\text{T}$ for a mass $m \sim 10^{10}$ amu (see~\ref{app:analysis_NV}). Satisfying this constraint, our scheme remains capable of probing MQC for such large masses, as illustrated by the following results of our analysis. \vspace{1mm}

\textit{Results for NV-center nanodiamond}: The required initial state of NV nanocrystal can be prepared by feedback cooling to ground state in low frequency traps such as ionic and diamagnetic traps \cite{nano_2016,PhysRevA.100.063819,nano_2020,Whittle2021GroundState10kg,Rudolph21_nanoparticle_cooling}. The embedded spin is then placed in a superposition $(\ket{+1}+\ket{-1})/\sqrt{2}$~\cite{Fang13_spinstate,Mamin14_spinstate,wood22_spinstate} and the nanocrystal is released from the trap into the confined magnetic field. Upon emerging, a spin readout can be implemented using established NV-center measurement techniques~\cite{measure_Hopper2018,measure_Holzgrafe2019,measure_Zhang2021,measure_Hensen2015,neumann_spin-readout_2010} to obtain the required spin statistics. For a readily achievable spin-coherence time $\tau = 1~\text{ms}$, we have identified the relevant $r$-regime as $10^{-5}<r<5\times 10^{-4}$ within which our coherence criterion is satisfied. Within this regime, our scheme can \textit{both} generate and certify MQC for spin-embedded nanodiamonds as massive as $10^{7}~\text{amu}$ using $B_0 = 5~m\text{T}$. Leveraging recent advances in NV electronic spin coherence under cryogenic conditions and operating within the regime where diamagnetic effects are negligible, we find that an extended coherence time of $\tau = 1~\text{s}$ ~\cite{1s_coherence_Abobeih2018,1s_coherence_PhysRevX.9.2019} would enable generation \textit{and} certification of MQC for masses approaching $10^{10}~\text{amu}$ for $B_0 = 5 ~\mu \text{T}$ (see~\ref{app:1s_coherence_NV}), with the delocalisation parameter reaching $d/\sigma = 4$ for $r=0.0004$ corresponding to $\mathcal{O} = 0.018$ \footnote{Note that a recent experiment \cite{SGQGtest1} based on full-loop SG interferometry has generated $d/\sigma \sim1.3$ for masses $\sim 100 ~\text{amu}$}. For instance, if $\sigma = 10~\text{nm}$ then we achieve a separation $d = 40~\text{nm}$. The key advantages of our scheme are that the spatial separation between the transmitted spatial components occurs along the direction of propagation and it does not change with time after emerging from the magnetic field region as long as coherence can be preserved. All these features reduce the experimental complexity in implementing our scheme. Note that, for a given mass, this spatial separation can be increased by decreasing the initial kinetic energy in a controlled manner for a given magnetic field. This provides an extra handle in comparison to SG-based schemes where, for a given mass, the superposition size can only be enhanced by appropriately controlling/enhancing the fields ~\cite{zhou2022,zhou2023,Zhou2024,Zhou2025,braccini2024}.

\textit{Outlook}--- In this paper, we have focused on both the generation and certification of spatial coherence via spin-only measurements using coherence parameter $\mathcal{O}$ and avoiding recombination, for which it has been necessary to restrict our attention to the “small-$r$” regime. We have also assumed conditions under which the diamagnetic effects are negligible. 
In the context of diamagnetism, note that as it is spin-independent, it does \textit{not} modify the basic spin-dependent splitting mechanism.
Further, in the sharp-field case, we have done a preliminary analysis (see~\ref{app:analysis_NV}) by considering an effective Hamiltonian where both spin-dependent potentials ($\pm \mu B_0$) are shifted by the spin independent diamagnetic term $\mathcal{D} =\frac{|\chi_m| M}{2\mu_0}\,{B_0}^2$ (where $M$ is the mass of the nanodiamond, $\chi_{m} = -6.2 \times 10^{-9} \,\, \mathrm{m}^3 \, \mathrm{Kg}^{-1}$ is the mass magnetic susceptibility, and $\mu_{0}$ is the vacuum permeability). Based on such a study, we have obtained a modified expression (see Eq.~\ref{eq:d_diamag} in \ref{app:analysis_NV}) for the spatial separation between the superposed states in terms of the parameters $r$ and $\eta = \mathcal{D}/\mathcal{Z}$ where $\mathcal{Z} = |\mu B_0|$.
It is then seen that for a spin coherence time of $\tau \approx 1s$ and $m=10^{10}$ amu, incorporating diamagnetism in our setup permits a higher $B_0 = 5~mT$, in place of the $5 ~\mu T$ required for assuming negligible diamagnetism, and enables generation of a superposition size $\sim 160 ~\mu m$ in the large-r regime for $r = 0.077$ (see~\ref{app:analysis_NV}). 
Thus, this preliminary study points to the promising role of diamagnetism in enabling access to higher field strengths and in enhancing the spatial superposition size (see Fig.~\ref{fig:d&d'_maintext}) in the large-r regime, while also ensuring $t_{f} < \tau$. Here we also consider the case of mass $m=10^{12}$ amu which is of particular interest in the context of proposed tabletop tests of quantum gravity \cite{sougato25_rev_mass}. Then, within a spin coherence time of $1s$ and $B_0 = 5~mT$, our setup can generate a spatial superposition of size $\sim 3 ~\mu m$ (see~\ref{app:analysis_NV}), which is already in the order of magnitude that suffices for gravitational entanglement when electromagnetic screening is used \cite{Schut2024micrometer}. Motivated by these encouraging observations, a worthwhile future direction would be to rigorously take into account the diamagnetic effects in our setup and comprehensively study different cases of superposition sizes and masses, taking cue from the existing studies~\cite{marshman2022}.


With respect to the "large-$r$" regime in our setup, we also note that, as $r \approx 1$ is approached, the kinetic energy of the incident particle becomes comparable to the Zeeman energy, and spatial component traversing the barrier is appreciably reflected, leading to a measurable reduction in the $+x$ spin probability of the transmitted wavefunction. This behavior can provide an alternative route (see~\ref{app:probs_analysis}), based on using $+x$ spin probability as a coherence parameter instead of $\mathcal{O}$, to discriminate spatially coherent states in this "large-$r$" regime from scenarios in which spatial coherence is absent.

Here we stress that our exact treatment of \textit{coupled} spatial--spin dynamics for particles traversing a confined uniform magnetic field has wider ramifications beyond the context of probing MQC. On one front, it naturally connects to the quantum clock framework pioneered by Büttiker~\cite{PhysRevB.27.6178}, where spin precession serves as a clock for the transit of a particle through a magnetic field region. Within our framework, the standard Larmor clock result is recovered in the $d/\sigma \rightarrow 0 $ limit, while our exact treatment quantifies corrections to this clock readout across the entire above-barrier regime ($r<1$). This, therefore, opens up promising directions for investigating its implications, particularly in relation to the transit time distributions~\cite{steiberg_rev,tunnelling_2017_jul,tunnelling_2017_dec,Ramos2020TunnellingTime,tunnelling_2026}. 

Furthermore, our scheme has direct relevance for weak-field magnetometry with high spatial resolution \cite{rondin2014_magnetometry,balasubramanian2008_magnetometry,B_sensing_90,B_sensing_07,B_sensing_08}. Firstly, in the kinetic regime, any semiclassical Larmor-based sensing schemes~\cite{B_sensing_08} would require corrections arising from the rigorous quantum treatment developed here. Additionally, the prepared spatial superposition in itself can provide an alternative sensing mechanism. This is because the details of the field profile are encoded in the separation $d$ between the $\psi^\pm_{T}$ components (see~\ref{app:sensing}). Since $d \propto w$, measuring smaller separations would enable resolving uniform magnetic fields over shorter length scales. This can have a range of applications, and for exploring how the intrinsic quantumness of such a sensor may provide potential metrological advantages ~\cite{RevModPhys.65.413_magneticfield,sage2013_magneticfield,Webb2021DiamondQuantumSensor,Hansen2023_magneticfield}.\\

\textit{Acknowledgments} - NS acknowledges support from the Department of Science and Technology (DST) through the INSPIRE-SHE programme, and thanks IISER, Mohali for facilitating this collaboration. NS also thanks Dr.~Debarshi Das and Prof.~Urbasi Sinha for valuable discussions during the QFTA 2025 conference. DH acknowledges support for completing this work as Distinguished Visitor to the Light and Matter Physics group, RRI, Bangalore under the National Quantum Mission of the DST, as well as support from NASI for initiating this research programme at Bose Institute, Kolkata. SB would like to acknowledge EPSRC grant EP/X009467/1 and STFC grant ST/W006227/1. SB’s contribution was made possible through the support of the WithOut SpaceTime (WOST) project, supported by Grant ID\#63683 from the John Templeton Foundation (JTF). SB’s research is funded by the Gordon and Betty Moore Foundation through Grant GBMF12328, DOI 10.37807/GBMF12328, and the Alfred P. Sloan Foundation under Grant No. G-2023-21130.

\bibliography{apssamp}

\onecolumngrid

\setcounter{section}{0}
\setcounter{equation}{0}
\setcounter{figure}{0}
\setcounter{table}{0}

\renewcommand{\theequation}{S.\arabic{equation}}
\renewcommand{\thefigure}{S.\arabic{figure}}
\renewcommand{\thetable}{S.\arabic{table}}
\renewcommand{\thesection}{S.\arabic{section}}
\newpage

\begin{center}
{\large\textbf{Supplemental Materials: ``\papertitle''}}
\end{center}

\setlength{\parindent}{0pt}

\section{Analysis for our spatial--spin coupled treatment} \label{app:analysis}

\subsection{Coherence parameter in terms of spin-statistics} \label{app:analysis_coh_param}

One of the central goals of this work is the certification of \textit{spatial} coherence via \textit{spin-only} measurements. This motivates the formulation of a coherence parameter that (i) distinguishes spin statistics arising from spatially coherent and incoherent states, and (ii) admits a direct relation to experimentally accessible spin observables. Thus, to arrive at a suitable coherence parameter we begin from the final spatial--spin coupled wavefunction from Eq.~(\ref{eq:state_at_tf}), which is of the form,
\begin{equation}
\Psi_{q}(\mathbf{x}, t_{f}) = \frac{1}{\sqrt{2}} \Big[ \psi^{+}\!(\mathbf{x},t_{f}) \otimes \ket{\uparrow}_{z} + \psi^{-}\!(\mathbf{x},t_{f}) \otimes \ket{\downarrow}_{z} \Big] \,.
\end{equation}
Considering the transmitted part of this wavefunction that emerges from the region of magnetic field, we get
\begin{equation}
\ket{\Psi_{q}} = \frac{1}{\sqrt{2}} \left( \ket{\psi^{+}_{T}} \otimes \ket{\uparrow}_z + \ket{\psi^{-}_{T}} \otimes \ket{\downarrow}_z \right) \,\,.
\end{equation}
The corresponding density matrix for this state is
\begin{equation}
\rho_{q} = \frac{1}{2} \left[ \ket{\psi^{+}_{T}}\bra{\psi^{+}_{T}} \otimes \ket{\uparrow}\bra{\uparrow} + \ket{\psi^{+}_{T}}\bra{\psi^{-}_{T}} \otimes \ket{\uparrow}\bra{\downarrow} + \ket{\psi^{-}_{T}}\bra{\psi^{+}_{T}} \otimes \ket{\downarrow}\bra{\uparrow} + \ket{\psi^{-}_{T}}\bra{\psi^{-}_{T}} \otimes \ket{\downarrow}\bra{\downarrow} \right] \,\,.
\end{equation}
Then, the reduced density matrix for the spin part can be written as follows,
\begin{equation}
\begin{aligned}
\rho_{q}^{\rm spin} &= \mathrm{Tr}_{\rm space}[\rho] \\
&= \frac{1}{2} \left[ T_{+} \, \ket{\uparrow}\bra{\uparrow} + \text{IP} \, \ket{\uparrow}\bra{\downarrow} + \text{IP}^{*} \, \ket{\downarrow}\bra{\uparrow} + T_{-} \, \ket{\downarrow}\bra{\downarrow} \right] \\
&= \frac{1}{2}
\begin{pmatrix}
T_{+} & \text{IP} \\
\text{IP}^{*} & T_{-}
\end{pmatrix} 
\,\,.
\end{aligned}
\end{equation}
Here, we define $\text{IP} = \braket{\psi^{+}_{T}}{\psi^{-}_{T}}$, $\text{IP}^{*} = \braket{\psi^{-}_{T}}{\psi^{+}_{T}}$, and $T_{\pm} = \braket{\psi^{\pm}_{T}}{\psi^{\pm}_{T}}$. \vspace{2mm}

Now, we can calculate the following spin expectation values as,
\begin{equation}
\begin{aligned}
&\langle \sigma_{x} \rangle = \mathrm{Tr} \left( \rho_{q}^{\rm spin} \, \sigma_{x} \right) = \frac{1}{2} \left( \text{IP} + \text{IP}^{*} \right) = \Re{\text{IP}} \,\,, \\
&\langle \sigma_{y} \rangle = \mathrm{Tr} \left( \rho_{q}^{\rm spin} \, \sigma_{y} \right) = \frac{i}{2} \left( \text{IP} - \text{IP}^{*} \right) = -\Im{\text{IP}} \,\,.
\end{aligned}
\end{equation}
Using these, we calculate the value of the squared inner product of the spatial components as,
\begin{equation}
\begin{aligned}
\abs{\braket{\psi^{+}_{T}}{\psi^{-}_{T}}}^{2} &= \left|\text{IP}\right|^{2} = \left|\text{IP}^{*}\right|^{2} \\
&= \left( \Re{\text{IP}} \right)^{2} + \left( \Im{\text{IP}} \right)^{2} \\
&= \langle \sigma_{x} \rangle^{2} + \langle \sigma_{y} \rangle^{2} \equiv \mathcal{O} \,\,.
\end{aligned}
\end{equation}
Motivated by the preceding analysis, we define $\mathcal{O} \equiv \expval{\sigma_x}^{2} + \expval{\sigma_y}^{2}$ as the \textit{coherence parameter} for certifying quantum spatial coherence within the framework developed in this paper. Furthermore, for the purposes of elegance and experimental simplicity, $\mathcal{O}$ is a good parameter as it comes out as a strictly monotonic function of the control parameter ($r$), unlike say spin probability along $+x$, or $\expval{\sigma_x}$, or $\expval{\sigma_y}$ which are periodic functions of $r$.

\subsection{Analytical solution for sharp potential} \label{app:analytical}
The "sharp" potential case refers to the scenario where the magnetic field rises/falls "sharply" at the boundaries. Such a magnetic field can be modeled as $B(x)\, \hat{z} = B_{0} \cdot f(x)\, \hat{z}$, where $f(x) =\frac{1}{2}\left[
\tanh\!\left(s\left(\frac{w}{2}+x\right)\right)
+
\tanh\!\left(s\left(\frac{w}{2}-x\right)\right)
\right]$, by taking $s$ to be large enough to approximate a sharp rise/fall (this has been done in the approximate solution and simulation for the smooth-potential case). In this section, however, we will stick to the ideal sharp potential given by
\begin{equation} \label{eq:analytic_intro1}
B(x) =
\begin{cases}
B_{0}, & |x| < \tfrac{w}{2}, \\[2pt]
0, & \text{elsewhere}.
\end{cases}
\end{equation}
Therefore, $B(x)\,\hat{z}$ essentially represents a uniform magnetic field of magnitude $B_{0}$ directed along the $+z$-axis within the region $|x|<w/2$.

\subsubsection{For plane wave} \label{app:analytical planewave}
We begin the analysis with an incident plane wave, $\varphi_{0} = A\,e^{ikx}$, with its spin polarized as $\ket{\chi} = \frac{1}{\sqrt{2}}(\ket{\uparrow}_z + \ket{\downarrow}_z)$. The complete incident wavefunction, therefore, is of the form $\ket{\Psi_{i}} = \varphi_{0} \otimes \ket{\chi}$. In our setup, the spin-down spatial component $(\varphi^{-})$ evolves through a potential well between $x = -w/2$ and $x = w/2$, and its reflected and transmitted components are respectively given by
\begin{equation} \label{eq:c1}
\begin{aligned}
\varphi_{R}^{-} &= A e^{-i k x} \frac{(k^{2} - k_{-}^{2})(1 - e^{2 i k_{-} w}) e^{-ikw}}{(k + k_{-})^{2} - (k - k_{-})^{2} e^{2 i k_{-} w}} \quad ; \quad x < -w/2 \\
\varphi_{T}^{-} &= A e^{i k x} \frac{4 k k_{-} e^{i k_{-} w} e^{-i k w}}{(k + k_{-})^{2} - (k - k_{-})^{2} e^{2 i k_{-} w}} \quad ; \quad x > w/2
\end{aligned}
\end{equation}
where, $w$ denotes the width of the region containing the uniform magnetic field, $k = \frac{\sqrt{2 m K}}{\hbar}$, and $k_{-} = \frac{\sqrt{2 m (K - \mu B_{0})}}{\hbar}$. Here, $K$ is the initial kinetic energy of the particle and $\mu$ is the magnetic moment. \vspace{3mm}

Similarly, the spin-up spatial component $(\varphi^{+})$ evolves under a potential barrier. Its reflected and transmitted parts respectively are
\begin{equation} \label{eq:c2}
\begin{aligned}
\varphi_{R}^{+} &= A e^{-i k x} \frac{(k^{2} - k_{+}^{2})(1 - e^{2 i k_{+} w}) e^{-ikw}}{(k + k_{+})^{2} - (k - k_{+})^{2} e^{2 i k_{+} w}} \quad ; \quad x < -w/2 \\
\varphi_{T}^{+} &= A e^{i k x} \frac{4 k k_{+} e^{i k_{+} w} e^{-i k w}}{(k + k_{+})^{2} - (k - k_{+})^{2} e^{2 i k_{+} w}} \quad ; \quad x > w/2
\end{aligned}
\end{equation}
with $k_{+} = \frac{\sqrt{2 m (K + \mu B_{0})}}{\hbar}$. \vspace{2mm}

In the regime where $K > |\mu B_{0}|$, the final state of the transmitted wavefunction can be recast in a compact form that explicitly reflects the spin rotation induced by the magnetic field. This expression represents a modified Larmor precession relation, derived from the time-evolved solutions of the spatial components $\varphi^{+}$ and $\varphi^{-}$,
\begin{equation} \label{eq:c3}
\ket{\Psi_{f}^{\prime}} = \frac{A e^{i k x}}{\sqrt{b^{2}+c^{2}}} \left( c \, e^{i \phi_{1}} \ket{\uparrow}_z + b \, e^{i \phi_{2}} \ket{\downarrow}_z \right)
\end{equation}
Here, the coefficients $c$ and $b$ represent the magnitudes of the transmitted spin-up and spin-down components, respectively, and are defined as
\begin{equation} \label{eq:c4}
\begin{aligned}
c &= \sqrt{ \left( \operatorname{Re} (\varphi_{T}^{+}) \right)^2 + \left( \operatorname{Im} (\varphi_{T}^{+}) \right)^2 } \\
b &= \sqrt{ \left( \operatorname{Re} (\varphi_{T}^{-}) \right)^2 + \left( \operatorname{Im} (\varphi_{T}^{-}) \right)^2 }
\end{aligned}
\end{equation}
The associated phases $\phi_1$ and $\phi_2$ are given by
\begin{equation} \label{eq:c5}
\begin{aligned}
\phi_{1} &= \tan^{-1} \left( \frac{ \operatorname{Im} (\varphi_{T}^{+}) }{ \operatorname{Re} (\varphi_{T}^{+}) } \right) \\
\phi_{2} &= \tan^{-1} \left( \frac{ \operatorname{Im} (\varphi_{T}^{-}) }{ \operatorname{Re} (\varphi_{T}^{-}) } \right)
\end{aligned}
\end{equation}
This analysis is in tune with the treatment given in~\cite{home2013} and the detailed expressions for $b$, $c$, $\phi_1$, and $\phi_2$ are available in the mentioned reference. Eq.~(\ref{eq:c3}) thus encapsulates the final state of the transmitted wavefunction and provides the modified Larmor precession phase, incorporating the influence of the spin-magnetic field interaction through the spatially resolved dynamics. \vspace{2mm}

Now, note that the standard semiclassical treatment of Larmor precession will yield the following emergent wavefunction~\cite{home2013}
\begin{equation} \label{eq:c6}
\ket{\Psi_{f}} = \frac{e^{i \phi/2}}{\sqrt{2}} \, \varphi_0 \otimes \left( \ket{\uparrow}_z + e^{-i \phi} \ket{\downarrow}_z \right)
\end{equation}
Here, $\phi = \omega T$ is the well-known Larmor precession phase, $\omega = \frac{2\mu B_{0}}{\hbar}$ is the  Larmor frequency, and $e^{i \phi/2}$ is a global phase factor (and hence can be ignored). In this context, the interaction time $T$ is identified with the traversal time of the particle through the magnetic-field region of width $w$, given by $T = \frac{w}{v}$, where the velocity $v$ is expressed as $v = \frac{\hbar k}{m}$ with $k$ being the initial wave number of the particle.

\subsubsection{For Gaussian wavepacket} \label{app:analytical gaussian}
The above results can then be extended for an incident Gaussian \textit{wavepacket} as the spatial component of the state entering the region of magnetic field, instead of a \textit{plane wave}. We consider an incident wave packet with its spin polarized as $\ket{\chi}$, and a magnetic field aligned along the $+z$-axis. The initial state is then given by  
\begin{equation} \label{eq:c7}
    \ket{\Psi_i} = \frac{1}{(2\pi\sigma^2)^{1/4}} e^{-(x - x_0)^2 / 4\sigma^2} e^{ik_0 x} \frac{1}{\sqrt{2}} (\ket{\uparrow}_z + \ket{\downarrow}_z) \,\, ,
\end{equation}
where $x_0$ is the wave packet’s center, $k_0$ is the central wave number, and $\sigma$ is its spatial width.

Since the spin precession induced by the magnetic field depends only on the wave number $k$, it is convenient to work in the momentum representation,
\begin{equation} \label{eq:c8}
\ket{\Psi_i} = \left(\frac{2\sigma^2}{\pi}\right)^{1/4} e^{-\sigma^2(k - k_0)^2} e^{ikx_0} \frac{1}{\sqrt{2}} (\ket{\uparrow}_z + \ket{\downarrow}_z) \,\, .
\end{equation}
From the Hamiltonian in Eq.~(\ref{eq:Pauli_eqs & Hamiltonian}), it follows that, due to the spin--magnetic field interaction, one spinor component evolves under an effective potential barrier while the other evolves under an effective potential well (depending on the sign of $\mu$). Consequently, the component experiencing the well propagates further, leading to a spatial separation between the two components along the propagation axis, which persists even after the spin-embedded system exits the magnetic-field region. Since the magnetic moments of neutrons and electrons are negative, we take $\mu<0$ in the analysis that follows. For systems with positive $\mu$, the dynamics of the spin-up and spin-down components are simply interchanged. Then, using the spin evolution results from Sec.~\ref{app:analytical planewave}, the final state becomes
\begin{equation} \label{eq:c9}
\ket{\Psi_{f}^{\prime}} = \left(\frac{2\sigma^2}{\pi}\right)^{1/4} e^{-\sigma^2(k - k_0)^2} \frac{1}{\sqrt{2}} \left( c(k) e^{ikx_{c}} e^{i\phi_1(k)} \, \ket{\uparrow}_z + b(k) e^{ik(x_{c}+d)} e^{i\phi_2(k)} \, \ket{\downarrow}_z \right) \,\, ,
\end{equation}
where $c(k)$, $b(k)$, $\phi_1(k)$, and $\phi_2(k)$ are the wavenumber-dependent quantities defined in Eqs.~(\ref{eq:c4})–(\ref{eq:c5}). This expression reveals that the spin–magnetic field interaction leads to a spin distribution over the different $k$-components of the wavepacket. One can then recast this final state in the position representation as
\begin{equation} \label{eq:c10}
\begin{aligned}
\ket{\Psi_{q}} = \ket{\Psi_{f}^{\prime}} &= \frac{1}{(2\pi\sigma^2)^{1/4}} \frac{1}{\sqrt{2}} \left( e^{-(x - x_{c})^2 / 4\sigma^2} \, \varphi^{+}_{T} \, \ket{\uparrow}_z + e^{-(x - (x_{c}+d))^2 / 4\sigma^2} \, \varphi^{-}_{T} \, \ket{\downarrow}_z \right) \\
&= \frac{1}{\sqrt{2}} \left( \ket{\psi^{+}_{T}} \otimes \ket{\uparrow}_z + \ket{\psi^{-}_{T}} \otimes \ket{\downarrow}_z \right) \,\, ,
\end{aligned}
\end{equation}
here, $\varphi^{\pm}_{T}$ are the plane-wave components associated with spin-up and spin-down, respectively, as obtained earlier in Eqs.~(\ref{eq:c1}) \&~(\ref{eq:c2}), and $\ket{\psi^{\pm}_{T}}$ represent the outgoing spatial wavepacket components. Note that $\ket{\psi^{+}_{T}}$ is centered at $x_c$, whereas $\ket{\psi^{-}_{T}}$ is centered at $x_{c}+d$; $d$ being the separation between the two peaks. The reduced density matrix for the spin part of the above state can be written as
\begin{equation}
\rho_{q}^{\rm spin} = \frac{1}{2} \left[ T_{+} \, \ket{\uparrow}\bra{\uparrow} + \text{IP} \, \ket{\uparrow}\bra{\downarrow} + \text{IP}^{*} \, \ket{\downarrow}\bra{\uparrow} + T_{-} \, \ket{\downarrow}\bra{\downarrow} \right] \,\,.
\end{equation}

\vspace{2mm} \noindent \underline{Computing explicit form of the parameter $\mathcal{O}$}: \vspace{1mm} \label{app:gaussian_coh_para,etc}

\noindent Note that,
\begin{equation} \label{eq:P'_parts_integrals}
\begin{aligned}
\text{IP} = \int_{\frac{w}{2}}^{\infty} \left(\psi_{T}^{+}(x)\right)^{*} \psi_{T}^{-}(x) \,\, dx \quad&,\quad \text{IP}^{*} = \int_{\frac{w}{2}}^{\infty} \left(\psi_{T}^{-}(x)\right)^{*} \psi_{T}^{+}(x) \,\, dx \,\,, \\
T_{+} = \int_{\frac{w}{2}}^{\infty} \left|\psi_{T}^{+}(x)\right|^{2} \, dx \quad&,\quad T_{-} = \int_{\frac{w}{2}}^{\infty} \left|\psi_{T}^{-}(x)\right|^{2} \, dx \,\,.
\end{aligned}
\end{equation}
Now, for simplicity we define the factors accompanying the transmitted states in Eqs.~\ref{eq:c1} \&~\ref{eq:c2} as follows
\begin{equation} \label{eq:factor_C+&C-}
\begin{aligned}
&C_{+} = \frac{4 k k_{+} e^{i k_{+} w} e^{-i k w}}{(k + k_{+})^{2} - (k - k_{+})^{2} e^{2 i k_{+} w}} \\
&C_{-} = \frac{4 k k_{-} e^{i k_{-} w} e^{-i k w}}{(k + k_{-})^{2} - (k - k_{-})^{2} e^{2 i k_{-} w}}
\end{aligned}
\end{equation}
Next, we can simplify and compute the integrals in Eq.~\ref{eq:P'_parts_integrals} to obtain,
\begin{equation} \label{eq:P'_parts_solved_integrals}
\begin{aligned}
&\text{IP} = \frac{(C_{+})^{*}C_{-}}{2} \, e^{-d^{2}/8\sigma^{2}} \, \textrm{erfc}\left(\frac{\frac{w}{2}-\left(x_{c}+\frac{d}{2}\right)}{\sqrt{2}\,\sigma}\right) \,\,, \\
&\text{IP}^{*} = \frac{(C_{-})^{*}C_{+}}{2} \, e^{-d^{2}/8\sigma^{2}} \, \textrm{erfc}\left(\frac{\frac{w}{2}-\left(x_{c}+\frac{d}{2}\right)}{\sqrt{2}\,\sigma}\right) \,\,, \\
&T_{+} = \frac{|C_{+}|^{2}}{2} \,\, \textrm{erfc}\left(\frac{\frac{w}{2}-x_{c}}{\sqrt{2}\,\sigma}\right) \,\,, \\
&T_{-} = \frac{|C_{-}|^{2}}{2} \,\, \textrm{erfc}\left(\frac{\frac{w}{2}-(x_{c}+d)}{\sqrt{2}\,\sigma}\right) \,\,.
\end{aligned}
\end{equation}
Here, the complementary error function, $\textrm{erfc}(x)$ is defined as $\textrm{erfc}(x) = 1 - \textrm{erf}(x) = \frac{2}{\sqrt{\pi}} \int_{x}^{\infty} e^{-t^{2}} dt$. 

Note that $x_{c} = \frac{w}{2} + \Delta x$, where $\Delta x$ denotes the distance between the edge of the region of magnetic field and the centre of the transmitted $\psi^{+}$ Gaussian component which is accounted for to ensure that a significant part of the component is outside the region of magnetic field when the spin measurement is carried out. Here, $\Delta x \sim n \sigma$, where $n$ is chosen suitably to ensure that a significant part of the transmitted component is out of the magnetic-field region. Under these circumstances, the $\textrm{erfc}(\,)$ factor effectively simplifies to
\begin{equation} \label{eq:erfc_term_simplified}
\textrm{erfc}\left(\frac{\frac{w}{2} - x_{c}}{\sqrt{2}\,\sigma}\right) = \textrm{erfc}\left(\frac{-n\,\sigma}{\sqrt{2}\,\sigma}\right) = \textrm{erfc}\left(\frac{-n}{\sqrt{2}}\right)
\end{equation}
For all reasonable choices of $n \geq 2$, the above expression will yield $\textrm{erfc}\left(\frac{-n}{\sqrt{2}}\right) \approx 2$. Similar reasoning holds for the other $\textrm{erf(\,)}$ terms as well. Using this simplification, we can write,
\begin{equation} \label{eq:simplified_params}
\begin{aligned}
&\text{IP} \approx \left[(C_{+})^{*}C_{-}\right] \, e^{-d^{2}/8\sigma^{2}} \,\, , \,\, \text{IP}^{*} \approx \left[(C_{-})^{*}C_{+}\right] \, e^{-d^{2}/8\sigma^{2}} \,\,, \\
&T_{+} \approx |C_{+}|^{2} \,\,, \text{ and } \,\, T_{-} \approx |C_{-}|^{2} \,\,.
\end{aligned}
\end{equation}
Then, the square of the overlap between the two spatial components is calculated as
\begin{equation} \label{eq:IPsq_spin-expectation}
\begin{aligned}
&\left|\text{IP}\right|^{2} = \left|\text{IP}^{*}\right|^{2} \\
&\approx |C_{+}|^{2} |C_{-}|^{2} \, e^{-d^{2}/4\sigma^{2}} \\
&\approx T_{+} \, T_{-} \, e^{-d^{2}/4\sigma^{2}} \,\,.
\end{aligned}
\end{equation}
Therefore, the parameter $\mathcal{O}$ has the following final form,
\begin{equation} \label{eq:coh_param_O_form}
\begin{aligned}
\mathcal{O} &= \langle \sigma_{x} \rangle^{2} + \langle \sigma_{y} \rangle^{2} \\
&= \frac{|C_{+}|^{2} |C_{-}|^{2}}{4} \, e^{-d^{2}/4\sigma^{2}} \, \left[\textrm{erfc}\left(\frac{\frac{w}{2}-\left(x_{c}+\frac{d}{2}\right)}{\sqrt{2}\,\sigma}\right)\right]^{2} \\
&\approx |C_{+}|^{2} |C_{-}|^{2} \, e^{-d^{2}/4\sigma^{2}} \,\,.
\end{aligned}
\end{equation}
Here, we calculate $|C_+|^{2}$ and $|C_-|^{2}$ using Eq.~\ref{eq:factor_C+&C-} as follows,
\begin{equation} \label{eq:C+/-_squared}
\begin{aligned}
&\left|C_{+}\right|^{2} = \frac{16 k^{2} k_{+}^{2}}{\left(k + k_{+}\right)^{4} + \left(k - k_{+}\right)^{4} - 2\left(k + k_{+}\right)^{2} \left(k - k_{+}\right)^{2} \cos\left(2k_{+}w\right)} \,\,, \\
&\left|C_{-}\right|^{2} = \frac{16 k^{2} k_{-}^{2}}{\left(k + k_{-}\right)^{4} + \left(k - k_{-}\right)^{4} - 2\left(k + k_{-}\right)^{2} \left(k - k_{-}\right)^{2} \cos\left(2k_{-}w\right)} \,\,.
\end{aligned}
\end{equation}

\underline{Size of the macroscopic superposition}: \vspace{1mm} \label{app:gaussian_separation}

In this context, note that the peaks of the two outgoing Gaussian spatial components in the quantum (spatial-spin coupled) treatment are separated by some distance $d$. This distance of separation between the components (\textit{i.e.}, the superposition size) can be obtained as follows.

The time evolution of the expectation value of the position operator $\hat{x}$ is governed by the equation of motion
\begin{equation} \label{eq:com_evol1}
\frac{d \langle\hat{x}\rangle}{dt} = \frac{i}{\hbar} \left\langle \left[\hat{H}, \hat{x}\right] \right\rangle \,.
\end{equation}
We consider, for instance, the trajectory of the wavepacket associated with the $\ket{\uparrow}_{z}$ component, denoted as $\psi^{+}$. An analogous analysis applies to $\psi^{-}$ upon employing the Hamiltonian $H_{-}$ in place of $H_{+}$. Then, by plugging in the operator equivalent of $H_{+}$ from Eq.~\ref{eq:Pauli_eqs & Hamiltonian} into Eq.~\ref{eq:com_evol1}, we obtain
\begin{equation} \label{eq:com_evol2}
\begin{aligned}
\frac{d \langle\hat{x}_{+}\rangle}{dt} &= \frac{i}{\hbar} \left[ \int \left(\psi^{+}(x)\right)^{*}\, \frac{\hbar^2}{2m}\left(x\frac{\partial^2}{\partial x^2}\left(\psi^{+}(x)\right) - \frac{\partial^2}{\partial x^2}\left(x\psi^{+}(x)\right)\right) \,dx \right] \\
&= \frac{i}{\hbar} \left[ -\frac{\hbar^{2}}{m} \int (\psi^{+}(x))^{*}\,\frac{\partial}{\partial x}\left(\psi^{+}(x)\right) \,dx \right] \\
&= \frac{i}{\hbar} \left[ -\frac{i\hbar}{m} \left\langle\hat{p}_{+}\right\rangle \right] \\
&= \frac{\left\langle\hat{p}_{+}\right\rangle}{m}
\end{aligned}
\end{equation}
Here, $\langle\hat{p}_{+}\rangle$ is the expectation value of the momentum operator corresponding to $\psi^{+}$. Similarly, for $\psi^{-}$, we will have $\frac{d \langle\hat{x}\rangle}{dt} = \frac{\left\langle\hat{p}_{-}\right\rangle}{m}$, where $\langle\hat{p}_{-}\rangle$ is the expectation value of the momentum operator corresponding to $\psi^{-}$.

Putting simply, $\Delta \langle\hat{x}\rangle = \langle\hat{p}_{\pm}\rangle\Delta t/m$, meaning that the average quantities $\langle\hat{x}\rangle$ and $\langle\hat{p}\rangle$ obey the classical relation $\Delta x = (p/m)\Delta t$ (a consequence of the Ehrenfest theorem~\cite{ehrenfest}). Consequently, the center-of-mass trajectory of each spatial component $\psi^{\pm}$ can be determined directly from the corresponding momentum and the elapsed time of evolution at each stage, up to their emergence from the region of uniform magnetic field. \vspace{2mm}

Against the above backdrop consider the following. Within the region of magnetic field, the $\psi^{-}$ component travels at $k_{-}$ and the $\psi^{+}$ component travels at $k_{+}$, with appropriate units of $\hbar$ and $m$. Here, $k_{-} = \frac{\sqrt{2m(K + |\mu B_{0}|)}}{\hbar}$, and $k_{+} = \frac{\sqrt{2m(K - |\mu B_{0}|)}}{\hbar}$; $K = \frac{\hbar^{2}k^{2}}{2 m}$. Clearly, $k_{-} > k_{+}$. The corresponding velocities are
\begin{equation} \label{eq:c_vels}
v_{-} = \frac{\hbar k_{-}}{m} = \sqrt{\frac{2(K + |\mu B_{0}|)}{m}}, \quad \text{and} \quad v_{+} = \frac{\hbar k_{+}}{m} = \sqrt{\frac{2(K - |\mu B_{0}|)}{m}} \,\, .
\end{equation}
Let $t_1$ be the time that the $\psi^{-}$ component takes to reach the boundary of the field at $x = w/2$. Then, $t_1 = w/v_{-}$. Let the separation developed between the two components till $\psi^{-}$ reaches the boundary be $d_1$. Then,
\begin{equation} \label{eq:c13}
d_{1} = w - v_{+}t_{1} = w\left( 1 - \sqrt{\frac{K - |\mu B_{0}|}{K + |\mu B_{0}|}} \right)
\end{equation}
Let $t_2 = w/v_{+}$ be the time that the $\psi^{+}$ component takes to reach the edge of the field at $x = w/2$. Then, the difference is $\Delta t = (t_2 - t_1)$. The additional separation developed between the two components till $\psi^{+}$ reaches the boundary is
\begin{equation} \label{eq:c14}
\begin{aligned}
d_{2} &= v \Delta t - v_{+} \Delta t \\
&= \left( \sqrt{\frac{2K}{m}} - \sqrt{\frac{2(K - |\mu B_{0}|)}{m}} \right) \times w \left( \sqrt{\frac{m}{2(K - |\mu B_{0}|)}} - \sqrt{\frac{m}{2(K + |\mu B_{0}|)}} \right) \\
&= w\left( \frac{(\sqrt{2K} - \sqrt{2(K - |\mu B_{0}|)}) (\sqrt{2(K + |\mu B_{0}|)} - \sqrt{2(K - |\mu B_{0}|)})}{\sqrt{2(K - |\mu B_{0}|)} \sqrt{2(K + |\mu B_{0}|)}} \right)
\end{aligned}
\end{equation}
We can substitute $|\mu B_{0}|/K = r$ in Eqs.~(\ref{eq:c13}) \&~(\ref{eq:c14}) to get
\begin{equation} \label{eq:c15}
\begin{aligned}
d_1 &= w \left( 1 - \sqrt{\frac{1-r}{1+r}} \right) \,\, , \\
d_2 &= w \left( \frac{(1 - \sqrt{1-r}) (\sqrt{1+r} - \sqrt{1-r})}{\sqrt{1+r}\sqrt{1-r}} \right) \,\, .
\end{aligned}
\end{equation}
Therefore, as $\psi^{+}$ emerges from the region of magnetic field, the total separation between the two spatial components $\psi^{+}$ \& $\psi^{-}$ is
\begin{equation} \label{eq:c16}
\begin{aligned}
d &= d_1 + d_2 \\
&= w \left( \frac{\sqrt{1+r} - \sqrt{1-r}}{\sqrt{1+r}} + \frac{(1 - \sqrt{1-r}) (\sqrt{1+r} - \sqrt{1-r})}{\sqrt{1+r}\sqrt{1-r}} \right) \\
&= w \left( \frac{1}{\sqrt{1-r}} - \frac{1}{\sqrt{1+r}} \right)
\end{aligned}
\end{equation}
Eq.~(\ref{eq:c16}) represents the final macroscopic separation between the spatial components of the emergent state. Furthermore, this macroscopic separation is intrinsically related to the coherence parameter, $\mathcal{O}$ as follows,
\begin{equation} \label{eq:IPsq_expression}
\mathcal{O} \approx T_{+} \, T_{-} \, e^{-d^{2}/4\sigma^{2}} \,\,.
\end{equation}
Therefore, we have the following relation,
\begin{equation} \label{eq:IPsq_T+_T-_d&sigma}
d^{2} \approx 4\sigma^{2} \, \ln{\left(\frac{T_{+}\,T_{-}}{\mathcal{O}}\right)}
\end{equation}

\subsubsection{Maximum mass and minimum velocity} \label{app:analytical constraints}
We begin by noting that $r = |\mu B_{0}|/K$, where $K = 3k_{B}T/2 = mv^{2}/2$.
\begin{equation} \label{eq:con1}
\implies v = \sqrt{\frac{2|\mu B_{0}|}{mr}}
\end{equation}
Now, note that the solutions detailed in Secs.~\ref{app:analytical planewave} \&~\ref{app:analytical gaussian} are valid in the regime $0<r<1$. This imposes a constraint on the velocity and the mass of the system ($v>\sqrt{\frac{2|\mu B_{0}|}{m}}$). Moreover, notice from Eq.~(\ref{eq:con1}) that as $m$ is increased, $v$ must decrease for fixed $\mu,\, B, \text{ and }\,r$. This is because fixing these parameters fixes the energy of the system (both $K$ and $\mu B$) and thus the particle is allowed to be more massive, so long as it is compensated by a lower velocity. \vspace{1mm}

Additionally, for a fixed barrier width $w$, the coherence time $\tau$ imposes a constraint on the total time of evolution $t_{f}$ given as
\begin{equation} \label{eq:con2}
t_{f} = \frac{w}{v_{+}} + \frac{\Delta x}{v} \leq \tau
\end{equation}
Here, while constructing the above inequality we recall that one spatial component leads the other as they emerge from the region of magnetic field. So, $w/v_{+}$ accounts for the time spent within this region by the slower spatial component ($\psi^{+}$). $\Delta x/v$ denotes the extra time accounted for the complete wavepacket to emerge out. Here, $\Delta x \sim n\sigma$, where $\sigma$ denotes the spread of the wavepacket in tune with earlier denotations and $n$ can be chosen suitaby to ensure that the complete wavepacket is out of the barrier. \vspace{2mm}

\noindent As stipulated earlier, $v_{+} = \frac{\hbar k_{+}}{m} = \sqrt{\frac{2(K - |\mu B_{0}|)}{m}} = \sqrt{\frac{2K(1-r)}{m}}$. On substituting, Eq.~(\ref{eq:con2}) reads,
\begin{equation} \label{eq:con3}
\begin{aligned}
&w\sqrt{\frac{m}{2K(1-r)}} + \frac{\Delta x}{v} \leq \tau \\
\Rightarrow \,\, &v \geq \frac{\Delta x}{\left(\tau - w\sqrt{\frac{m}{2K(1-r)}}\right)} \\
\Rightarrow \,\, &v \geq \left(\frac{\Delta x \sqrt{2K(1-r)}}{\tau \sqrt{2K(1-r)} - w\sqrt{m}}\right)
\end{aligned}
\end{equation}
Next, recall that we deduced from Eq.~(\ref{eq:con1}) that the particle's initial velocity can be lowered if that is compensated by increasing the mass proportionally. However, this lowering of velocity cannot be beyond the constraint imposed by the coherence time of the system. Therefore, the maximum mass allowed for a set of parameters corresponds to the situation where these two constraints match. Thus,
\begin{equation} \label{eq:con4}
\begin{aligned}
\sqrt{\frac{2|\mu B_{0}|}{mr}} = \frac{\Delta x \sqrt{2K(1-r)}}{\tau \sqrt{2K(1-r)} - w\sqrt{m}} &\,\,\Rightarrow \sqrt{\frac{2K}{m}} = \frac{\Delta x \sqrt{2K}\sqrt{(1-r)}}{\tau \sqrt{2K(1-r)} - w\sqrt{m}} \\
&\,\,\Rightarrow \sqrt{m} \left( \Delta x \sqrt{(1-r)} + w \right) = \tau \sqrt{2K(1-r)} \\
&\,\,\Rightarrow \sqrt{m} = \frac{\tau \sqrt{2|\mu B_{0}|\left(\frac{1-r}{r}\right)}}{\Delta x \sqrt{(1-r)} + w}
\end{aligned}
\end{equation}
\begin{equation} \label{eq:con5}
\hspace{-1cm}\implies m_{\textrm{max}} = \frac{2\tau^{2} |\mu B_{0}|}{r \left( \Delta x + \frac{w}{\sqrt{1-r}} \right)^{2}} = \frac{2\tau^{2} K}{\left( \Delta x + \frac{w}{\sqrt{1-r}} \right)^{2}}
\end{equation}
The above expression for $m_{\textrm{max}}$ is the maximum permissible mass of the system in terms of the various key parameters like $\tau$, $B_{0}$, $w$, etc. Since, $|\mu B_{0}|/K = r$, we must have $m_{\textrm{max}} = 2|\mu B_{0}|/rv_{\textrm{min}}^{2}$. Then, the minimum initial velocity of the particles must be
\begin{equation} \label{eq:con6}
v_{\textrm{min}} = \frac{1}{\tau} \left( \Delta x + \frac{w}{\sqrt{1-r}} \right)
\end{equation}
Eqs.~(\ref{eq:con5}) \&~(\ref{eq:con6}) succinctly express the permitted maximum mass and minimum initial velocity of the system, respectively. In tune with earlier discussions, the regime of $r$ remains $0<r<1$ for these expressions.


\subsection{Approximate solution for smooth potential} \label{app:approx}

In the lab, uniform magnetic fields that are confined within a region tend to rise and fall smoothly at their edges unlike a sharp barrier or well. The ``smooth'' potential case refers to this situation where the magnetic field at the boundaries does not change abruptly, but rather varies smoothly with position. We model it by-
\begin{equation}
    \textbf{B}(x) = B_0\, f(x) \hat{z} \quad \text{where} \quad
    f(x) = \frac{1}{2}\left[
\tanh\!\left(s\left(\frac{w}{2}+x\right)\right)
+
\tanh\!\left(s\left(\frac{w}{2}-x\right)\right)
\right]
\end{equation}
Here $f(x)$ is a dimensionless function forming a smooth-edged barrier centered at $x=0$ of an approximate width 'w'. For experimental relevance, it is useful to quantify the smoothness of the transition at the barrier edges. A natural measure for this purpose is the spatial interval ($\Delta x_{\text{edge}}$) over which $B(x)$ rises from $0.1B_{0}$ to $0.9B_{0}$. For our potential profile one finds $\Delta x_{\text{edge}} \sim \mathcal{O}(1/s)$. Thus, the parameter $s$ sets an effective edge length scale $\ell_{\text{edge}}=1/s$: larger $s$ corresponds to sharper, more abrupt edges, while smaller $s$ yields a smoother ramp. Unlike the sharp step function considered in~\ref{app:analytical}, this case does not yield simple analytical solutions. Hence, we employ the WKB approximation to capture the effect of the slowly varying potential profile on the incident Gaussian wavepacket.

\subsubsection{WKB phase evolution} \label{app:approx wkb}

The stationary state Schrödinger equation along the direction of propagation (+x-axis) reads
\begin{equation} \label{eq:a6}
-\frac{\hbar^2}{2m}\psi''(x) + V(x)\psi(x) = E\psi(x) \, .
\end{equation}
For our slowly varying potential $V(x)$, we employ the following ansatz
\begin{equation} \label{eq:a7}
\psi(x) = A(x)\, e^{i\delta(x)} e^{ikx} = A(x) e^{i\varphi(x)}, 
\qquad \varphi(x) = kx + \delta(x),
\end{equation}
we assume $A(x)$ varies slowly. This ansatz is chosen to facilitate the analysis of the wavefunction sufficiently far away from the barrier where $V(x) = 0$, hence in the asymptotic region, we expect to see a plane wave solution. Differentiation yields
\begin{equation} \label{eq:a8}
\psi'(x) = e^{i\varphi(x)} \big(A' + i A \varphi' \big),
\end{equation}
\begin{equation} \label{eq:a9}
\psi''(x) = e^{i\varphi(x)} \big(A'' + 2i A'\varphi' + i A \varphi'' - A(\varphi')^2 \big).
\end{equation}
Substituting the above into Eq.~(\ref{eq:a6}) gives
\begin{equation} \label{eq:a10}
-\frac{\hbar^2}{2m} \big(A'' + 2i A'\varphi' + i A\varphi'' - A(\varphi')^2 \big) 
+ V A = E A \, .
\end{equation}
Separating real and imaginary parts,
\begin{equation} \label{eq:a11}
\frac{\hbar^2}{2m} (\varphi')^2 = E - V(x) - \frac{\hbar^2}{2m}\frac{A''}{A},
\end{equation}
\begin{equation} \label{eq:a12}
2A'\varphi' + A\varphi'' = 0 \, \implies A^2 \varphi' = C^2 \implies A = \frac{C}{\sqrt{|\varphi'(x)|}}
\end{equation}
Neglecting $A''/A$ (WKB approximation), Eq.~(\ref{eq:a11}) reduces to
\begin{equation} \label{eq:a13}
\frac{\hbar^2}{2m} (\varphi')^2 = E - V(x).
\end{equation}
and by defining the local wavenumber as
\begin{equation} \label{eq:a14}
k(x) = \frac{1}{\hbar}\sqrt{2m(E - V(x))} 
= \sqrt{k^2 - \alpha(x)}, 
\qquad \alpha(x) = \tfrac{2m}{\hbar^2} V(x),
\end{equation}
we obtain
\begin{equation} \label{eq:a15}
\varphi'(x) = k(x).
\end{equation}
Since $\varphi'(x) = k + \delta'(x)$, the phase shift relative to free propagation is
\begin{gather} \label{eq:a16}
\delta'(x) = k(x) - k,\\
\delta(k) = \int^x \big(k(x') - k\big) \, dx' \label{eq:a17}
\end{gather}
Equation~(\ref{eq:a17}) thus gives the accumulated WKB phase due to the smooth potential. Define the transmission amplitude $t(k) \equiv e^{i \delta(k)}$ so that the final solution takes the form 
\begin{equation}
    \psi(x) \approx \frac{C}{\sqrt{k(x)}}\, t(k) e^{ikx} = \frac{C}{\sqrt{k}}\, t(k) e^{ikx} 
    \quad (\,k(x) \rightarrow k \,\, \text{as} \,\, x \rightarrow \infty \,)
\end{equation}
Note that in deriving this result we have neglected reflection from the barrier. This is justified for the parameter regime relevant to generating and certifying macroscopic quantum coherence, namely the small-$r$ regime where $K \gg \mu B_0$ and transmission dominates. Furthermore, even in the large-$r$ regime, above-barrier reflection for smooth potentials is expected to be exponentially suppressed relative to the sharp-barrier case due to the smooth-edged nature of the potential~\cite{Jaffe2010MomentumSpaceTunneling}. This approximation is expected to break down only near $r \sim 1$ (i.e. $K \sim \mu B_0$), where reflection effects can no longer be neglected.

\subsubsection{Solution for Plane Wave}

Now we apply the method from the previous section (ref~\ref{app:approx wkb}) to solve for the spinor dynamics across the smooth potential region. We begin the analysis with an incident plane wave, $A\,e^{ikx}$, such that the complete incident wavefunction is of the form $\ket{\Psi(x,0)} = A\,e^{ikx} \, \otimes \, \frac{1}{\sqrt{2}}(\ket{\uparrow}_z + \ket{\downarrow}_z)$. Then we take into account the effect of the WKB phase on each spin component. From Eq.~(\ref{eq:a17}), the WKB phase acquired by spin up and spin down component is
\begin{equation} \label{eq:a20}
\delta_{\pm}(k) = \int^x \Bigg( \sqrt{k^2 - \frac{2m}{\hbar^2}V_{\pm}(x')} - k \Bigg) dx' .
\end{equation}
Thus, at the asymptotic region (sufficiently far from the field), the accumulated spin-dependent phases for both spinor components are - 
\begin{equation} \label{eq:a21}
\delta_\pm = \int_{-\infty}^{\infty} \left( \sqrt{k^2 \pm \frac{2m}{\hbar^2}\mu B(x)} - k \right) dx ,
\end{equation}
This provides us with the final transmitted wavefunction at time $t_{f}$ in the asymptotic region - 
\begin{equation} \label{eq:a22}
\ket{\Psi(x,t_f)} \approx \frac{Ce^{i(kx- \frac{\hbar k^{2}}{2m}t_f )}}{\sqrt{2k}}(t_{+}(k)\ket{\uparrow}_z + t_{-}(k)\ket{\downarrow}_z)
\end{equation}
Here, the relative phase shift between spin-up and spin-down is therefore given by
\begin{equation} \label{eq:a24}
\phi = \delta_- - \delta_+
\end{equation}
Note that this relative phase is directly analogous to the spin precession angle in the sharp-boundary model (see~\ref{app:analytical}). However, unlike the sharp case, the precession angle now depends on the detailed spatial profile of the magnetic field $B(x)$.

\subsubsection{Solution for Gaussian Wave Packet} \label{app:approx gaussian}

Beginning with the initial normalized Gaussian envelope with RMS width $\sigma$ given by
\begin{equation} \label{eq:a1}
\psi_i(x) = \frac{1}{(2\pi\sigma^2)^{1/4}} \exp\!\left(-\frac{(x-x_0)^2}{4\sigma^2}\right) e^{i k_0 x},
\end{equation}
satisfying
\begin{equation} \label{eq:a2}
\int_{-\infty}^{\infty} |\psi_i(x)|^2 \, dx = 1 \, .
\end{equation}
The corresponding spinor, initially prepared in an equal superposition in the $z$-basis, is
\begin{equation} \label{eq:a3}
\Psi(x,0) =\psi_i(x) \otimes \frac{1}{\sqrt{2}}
\begin{pmatrix}
1 \\[3pt]
1
\end{pmatrix}.
\end{equation}
Passing to momentum space, the Fourier transform of $\psi_i(x)$ is
\begin{equation} \label{eq:a4}
c(k) = \frac{1}{\sqrt{2\pi}} \int_{-\infty}^{\infty} \psi_i(x) \, e^{-ikx} \, dx \, ,
\end{equation}
which evaluates to the Gaussian profile
\begin{equation} \label{eq:a5}
c(k) = \left(\frac{2\sigma^2}{\pi}\right)^{1/4}
\exp\!\left(-\sigma^2 (k-k_0)^2 - i(k-k_0)x_0\right)
\end{equation}
Note that each plane wave $e^{ikx}$ component acquires a transmission amplitude ($t_\pm(k)$) after crossing the smooth potential:
\begin{equation}\label{eq:scattering_phase_smooth}
t_\pm(k) \approx e^{i\delta_\pm(k)}, \qquad
\delta_\pm(k) = \int_{-\infty}^{\infty}\Big(\sqrt{k^2-\alpha_\pm(x)}-k\Big)\,dx,
\end{equation}
with
\begin{equation}
\alpha_\pm(x)=\frac{2m}{\hbar^2}V_\pm(x),\qquad V_\pm(x)=\mp\mu B(x)
\end{equation}
Including free time evolution with $\omega(k)=\hbar k^2/(2m)$, the transmitted components are
\begin{equation} \label{eq:transmitted_integral}
\psi^{\pm}(x,t) = \int_{-\infty}^{\infty} c(k)\,\frac{e^{i\delta_\pm(k)}}{\sqrt{k}}\,e^{i(kx-\omega(k)t)}\,dk.
\end{equation}
Taking $c(k)$ to be sharply peaked at $k=k_0$, we expand the total phase of the integrand in Eq.~(\ref{eq:transmitted_integral}) about $k_0$.
Define $\Delta k = k - k_0$, and write the total phase as
\begin{equation}
\Phi_\pm(k;x,t) = kx - \omega(k)t + \delta_\pm(k) - (k-k_0)x_0 ,
\end{equation}
where the last term arises from the exponential factor in $c(k)$.
The integrand of Eq.~(\ref{eq:transmitted_integral}) can then be written as
\begin{equation}
c(k)\,e^{i\delta_\pm(k)}e^{i(kx-\omega t)}
= \tilde{A}\,e^{-\sigma^2 \Delta k^2}\,e^{i\Phi_\pm(k;x,t)}, \qquad
\tilde{A}=\left(\frac{2\sigma^2}{\pi k_0^2}\right)^{1/4}.
\end{equation}
Expanding $\omega(k)$ and $\delta_\pm(k)$ to second order around $k_0$ gives
\begin{align}
\omega(k) &= \omega_0 + \omega'_0 \Delta k + \tfrac{1}{2}\omega''_0 \Delta k^2,\\
\delta_\pm(k) &= \delta_{\pm,0} + \delta'_{\pm,0} \Delta k + \tfrac{1}{2}\delta''_{\pm,0} \Delta k^2,
\end{align}
where primes denote derivatives with respect to $k$ evaluated at $k_0$.
Substituting these expansions into $\Phi_\pm$ and collecting terms yields
\begin{equation}
\Phi_\pm = Z_\pm + L_{\pm}\,\Delta k + Q_{\pm}\,\Delta k^2,
\end{equation}
with
\begin{align}
Z_\pm &= k_0x - w_0t + \delta_{\pm}(k_0) , \\[4pt] 
L_{\pm} &= (x - x_0) - \omega'_0 t + \delta'_{\pm}(k_0), \label{eq:L}\\[4pt]
Q_{\pm} &= -\tfrac{1}{2}\omega''_0 t + \tfrac{1}{2}\delta''_{\pm}(k_0). \label{eq:Q}
\end{align}
The integral in Eq.~(\ref{eq:transmitted_integral}) is now Gaussian in $\Delta k$ and takes the form:
\begin{equation}\label{eq:approx_integralQ}
\tilde{A}e^{iZ_\pm}\int_{-\infty}^{\infty} 
\exp\!\big[-(\sigma^2 - iQ_\pm)\Delta k^2 + iL_\pm\,\Delta k \big]\, d\Delta k
= \tilde{A}e^{iZ_\pm}\sqrt{\frac{\pi}{\sigma^2 - iQ_\pm}}\,
\exp\!\left[-\frac{L_{\pm}^2}{4(\sigma^2 - iQ_{\pm})}\right].
\end{equation}
In the narrowband limit $|Q|\ll\sigma^2$, where higher-order dispersion is negligible, (for suitable parameter regimes one can include the dispersion term $Q\approx -\tfrac{1}{2}\omega''_0 t$) this reduces to
\begin{equation}
\psi^{\pm}(x,t) \propto e^{i(k_0x-\omega_0 t+ \delta_\pm(k_0))} 
e^{-\frac{(x - x_{\pm}(t))^2}{4\sigma^2}},
\end{equation}
Normalizing this solution provides the final transmitted wavefunction after time t - 
\begin{equation}\label{eq:wkb_soln}
\begin{split}
    \ket{\Psi(x,t)} &= \frac{1}{\sqrt{2}}(\psi^{+}(x,t)\ket{\uparrow}+\psi^{-}(x,t)\ket{\downarrow}) \\ \psi^{\pm}(x,t) &= (2\pi\sigma^2)^{-\frac{1}{4}} e^{i(k_0x-\omega_0 t+ \delta_\pm(k_0))} e^{-\frac{(x - x_{\pm}(t))^2}{4\sigma^2}}
\end{split}
\end{equation}
The transmitted packet thus retains its RMS width $\sigma$, while its centre follows
\begin{equation}
x_\pm(t) = x_0 + \omega'_0 t - \delta'_{\pm}(k_0), \qquad
\omega'_0 = \frac{d\omega}{dk}\Big|_{k_0} = \frac{\hbar k_0}{m} \equiv v_g.
\end{equation}
The term $\delta'_{\pm,0}$ therefore represents a scattering-induced group delay, corresponding to a spatial shift of $-\delta'_{\pm,0}$ in the wavepacket centre.
The differential displacement between the peaks of the gaussian wavepackets corresponding to the two spin components is
\begin{equation}
d= \Delta x = x_-(t) - x_+(t)
= -\delta'_-(k_0) + \delta'_+(k_0),
\end{equation}
which also indicates the reduction of their mutual overlap.
Within the WKB approximation, the derivative of the scattering phase can be written as
\begin{equation}
\frac{d\delta_\pm}{dk}
= \int_{-\infty}^\infty\!\left(\frac{k}{\sqrt{k^2-\alpha_\pm(x)}}-1\right)\!dx,
\qquad
\alpha_\pm(x) = \frac{2mV_\pm(x)}{\hbar^2},
\end{equation}
providing the numerical expression used to evaluate the group delay. In the sharp case (large $s$ limit), this reduces to:
\begin{equation}\label{eq:d_wkb}
\begin{split}
    d &= \int_{-\infty}^{\infty}\!\left(\frac{k_0}{\sqrt{k_0^2-\alpha_+}}-\frac{k_0}{\sqrt{k_0^2-\alpha_-}} \right)\! dx = \left(\frac{1}{\sqrt{1+2m\mu B_0/\hbar^2 k_{0}^2}}-\frac{1}{\sqrt{1-2m\mu B_0/\hbar^2 k_{0}^2}} \right)\ \int_{-w/2}^{w/2} dx \\\\ &=\left( \frac{1}{\sqrt{1-|\mu B_0|/K}} - \frac{1}{\sqrt{1+|\mu B_0|/K}} \right) w \implies d = w \left( \frac{1}{\sqrt{1-r}} - \frac{1}{\sqrt{1+r}} \right)
\end{split}
\end{equation}
which agrees with Eq.~\ref{eq:c16}.

The inner product of the two spinor components can be calculated as follows using the explicit form from Eq.~\ref{eq:wkb_soln}, and evaluating the Gaussian integral- 
\begin{align}
\langle \psi^{+} | \psi^{-} \rangle
=
\int_{-\infty}^{\infty} dx\,
(\psi^{+}(x,t))^*\,\psi^{-}(x,t) = e^{i\phi_0}\exp\!\left(-\frac{d^2}{8\sigma^2}\right) .
\end{align}
\noindent where $\phi_0 = \delta_-(k_0) - \delta_+(k_0)$. Thus, the coherence parameter comes out as- 
\begin{align}
\mathcal{O} = \abs{\braket{\psi^{+}}{\psi^{-}}}^2  = \exp\!\left(-\frac{d^2}{4\sigma^2}\right)
\end{align}

We can also calculate the probability of spin outcome $\ket{\chi}$ by projecting onto $\ket{\chi} = (\ket{\uparrow}_z + \ket{\downarrow}_z)/\sqrt{2}$ as done in eq.~\ref{eq:P'_parts}:
\begin{equation} 
P_{\chi}^{\prime} = \int \left| \langle{\chi |\Psi_{q}} \rangle \right|^{2} \, dx = \frac14\left( T_{+} + T_{-} \right) + \frac14 \left( \text{IP} + \text{IP}^{*}\right) = \frac14\left( T_{+} + T_{-} \right) + \frac12\,\mathrm{Re}\,\langle \psi^{+} | \psi^{-} \rangle
\end{equation}
and in the case where $K > \mu B$ when reflection is almost negligible and there is complete transmission from both spin components ($T_+=T_-=1$) leading to 
\begin{align}
P_{\chi}^{\prime}
= \frac{1}{2} + \frac{1}{2} \Re S =\frac{1}{2}+\frac{1}{2}\exp\!\Big(-\frac{d^2}{8\sigma^2}\Big)\cos(\phi_0).
\end{align}
For large separation $d\gg\sigma$, the interference term ($cos(\phi_0)$) vanishes and $P_{\chi}^{\prime} \to 1/2$. 

\subsubsection{Limiting Cases}\label{app:smooth_limiting}
\noindent\textit{Recovery of semiclassical Larmor phase}: \vspace{1mm} \\
Our defined small r regime, provided that $\sigma \ll w$, corresponds to the high kinetic energy limit ($\abs{\mu B_0} \ll K$). In this limit the square root in 
$\delta_\pm$ (Eq.~(\ref{eq:scattering_phase_smooth})) can be expanded to leading order:
\begin{equation}
\sqrt{k^2 - \alpha_{\pm}(x)} \approx k - \frac{\alpha_\pm(x)}{2k} 
= k \mp \frac{m\mu B(x)}{\hbar^2 k},
\end{equation}
so the scattering phases become
\begin{equation}
\delta_\pm(k_0) \approx \mp\frac{m\mu}{\hbar^2 k_0}
\int_{-\infty}^{+\infty}\!B(x)\,dx,
\end{equation}
giving a relative phase
\begin{equation}
\phi_0 = \delta_-(k_0) - \delta_+(k_0) 
\approx \frac{2m\mu}{\hbar^2 k_0}\int_{-\infty}^{+\infty}\!B(x)\,dx 
= \frac{2\mu B_0 w}{\hbar v_g} = \omega_L \tau,
\end{equation}
where $\omega_L = 2\mu B_0/\hbar$ is the Larmor frequency and 
$\tau = w/v_g$ the traversal time. Note that in the small r limit, the expression derived for  $\phi_0$ is independent of the parameter '$s$' which controls the edge smoothness of the magnetic field. \vspace{3mm}

\noindent\textit{Independence of $d$ from edge sharpness 
parameter $s$ at small $r$}: \vspace{1mm} \\
The spatial separation $d$ is governed by the group-delay integrals
\begin{equation}
\delta'_\pm(k_0) = \int_{-\infty}^{+\infty}\!
\left(\frac{k_0}{\sqrt{k_0^2 - \alpha_\pm(x)}} - 1\right)dx, 
\qquad \alpha_\pm(x) = \frac{2mV_\pm(x)}{\hbar^2},
\end{equation}
with $d = \delta'_+(k_0) -\delta'_-(k_0)$. In the small r regime, expanding the integrand to leading 
order in $\alpha_\pm/k_0^2$ gives:
\begin{equation}
\frac{k_0}{\sqrt{k_0^2 - \alpha_\pm}} - 1 
\approx \frac{\alpha_\pm}{2k_0^2} 
= \frac{mV_\pm}{\hbar^2 k_0^2},
\end{equation}
so that
\begin{align}
    d &= \delta'_+(k_0) -\delta'_-(k_0) \approx \frac{2m\mu}{\hbar^2 k_0^2}
    \int_{-\infty}^{+\infty}\!B(x)\,dx\\
    \implies d &\approx \frac{\mu B_0}{\hbar^2 k_0^2/2m}\int_{-\infty}^{+\infty}\!
    \frac{1}{2}\left[\tanh\!\left(s\!\left(\tfrac{w}{2}+x\right)\right) 
    + \tanh\!\left(s\!\left(\tfrac{w}{2}-x\right)\right)\right]dx = r w,
\end{align}
Consequently, in the small-$r$ regime, the observables $d ~(\approx rw)$, $\mathcal{O} ~(\approx e^{-r^2 w^2/ 4 \sigma^2})$, and $P'_\chi$ are insensitive to the edge sharpness of the field profile. \\
Furthermore, in the lower end of this small r regime ($d/\sigma \lesssim 0.2$), as $ d/\sigma \rightarrow 0$, the expression of $P'_\chi$ simplifies to 
\begin{equation}
P'_\chi \to \frac{1}{2} + \frac{1}{2}\cos(\omega_L\tau) = \cos^2\!\left(\frac{\omega_L\tau}{2}\right) = P_\chi,
\end{equation}
thereby reproducing the semiclassical Larmor result.

\subsubsection{Approximations and validity}

The following assumptions underlie our analysis: Firstly, the WKB approximation is valid, requiring the local de Broglie wavelength to be much smaller than the characteristic length scale of the barrier, or equivalently
\[
\left|\frac{1}{k(x)}\frac{dk(x)}{dx}\right| \ll k(x).
\]
Second, the incident wavepacket has a narrow momentum distribution, with a Gaussian amplitude \(c(k)\) sharply peaked around \(k_0\) and width \(\sigma_k = 1/(2\sigma) \ll k_0\). Finally, wavepacket spreading is negligible in the relevant regimes: for large masses \(m \gg 10^5\,\mathrm{amu}\), the quantity \(Q = -\tfrac{1}{2}\omega'' t + \tfrac{1}{2}\delta''(k_0)\) satisfies \(Q \ll \sigma^2\), while for lighter masses \(m \le 10^5\,\mathrm{amu}\) one has \(Q \sim \sigma\) but \(\delta''(k_0) \ll \omega''_0 t\), allowing us to retain only the term \(Q \approx -\tfrac{1}{2}\omega'' t\) in Eq.~\ref{eq:approx_integralQ}.

\newpage
\section{Treatments with no spatial coherence} \label{app:incoherent_models}

\subsection{Semiclassical treatment of Larmor Precession} \label{app:semiclassical}
In deriving the solution to the semiclassical case ($|\Psi_{sc} \rangle$), we consider a spin- 1/2 embedded system passing through a uniform magnetic field region of width w with the spin part evolving solely under the spin-magnetic field interaction (and gaining a relative phase \textit{i.e.} Larmor phase) independently from the spatial part which undergoes free evolution. 
The study of this case is of interest for two reasons - (i) It helps elucidate the limiting case, where kinetic energy significantly dominates over magnetic potential energy of our coupled treatment, thus the potential interaction does not significantly alter the momentum of the incident wavepacket components. (ii) Although a spatially localized magnetic field generically entails spatial-spin coupling within standard quantum mechanics, increasing mass may drive the spatial \textsl{dof} towards effectively classical behavior, suppressing coherent spatial superposition. In such a scenario, the spin would continue to evolve quantum mechanically under the magnetic interaction, while the spatial mode would propagate classically, precisely the semiclassical limit described earlier. Therefore, experimental validation of spatial coherence requires ruling out this semiclassical explanation.

Under the semiclassical approximation we start with the Hamiltonian
\begin{equation}
\hat{H}_{\pm} = \frac{\hat{p}^2}{2m} + V_{\pm}(x)
= -\frac{\hbar^2}{2m}\nabla^2 + V_{\pm}(x).
\end{equation}
The Schrödinger equation reads
\begin{equation}
\hat{H}_{\pm}\psi^{\pm}(x,t) = i\hbar \,\partial_t \psi^{\pm}(x,t).
\end{equation}
Using the ansatz (assuming separable evolution):
\begin{equation}
\psi^{\pm}(x,t) = \psi_{\text{free}}(x,t)\, e^{i\theta_{\pm}(t)},
\end{equation}
and invoking
\begin{equation}
\frac{\hat{p}^2}{2m}\psi_{\text{free}} = i\hbar \,\partial_t \psi_{\text{free}},
\end{equation}
we obtain
\begin{equation}
\hbar \dot{\theta}_{\pm}(t) = - V_{\pm}(x).
\end{equation}
Assuming $\theta=\theta(t)$ and constant velocity $v=\hbar k_0/m$ (If the potential exerts a force, i.e. nonzero spatial gradient, the velocity becomes position-dependent: $ m \dot{v} = -\partial_x V_{\pm}(x)$. In that case one must use the actual classical trajectory $x(t)$ and integrate $\int V(x(t)) \, dt$), one finds
\begin{equation}
\frac{d\theta_{\pm}}{dx} = \frac{-V_{\pm}(x)}{\hbar v}, 
\qquad
\theta_{\pm} = \frac{-1}{\hbar v} \int_{x_i}^{x_f} V_{\pm}(x)\, dx.
\end{equation}
The relative phase is
\begin{equation}
\phi = \theta_{+}-\theta_{-} 
= \frac{2}{\hbar v} \int V(x)\, dx. \label{eq:Theta}
\end{equation}

For a square barrier
\begin{equation}
V(x) = 
\begin{cases}
0, & |x|>w/2, \\
\mu B, & |x|<w/2,
\end{cases}
\end{equation}

the relative phase is  
\begin{equation}
\phi = \frac{2}{\hbar v} \int V(x)\, dx
= \frac{2}{\hbar v} \mu B w.
\end{equation}
which matches with the textbook Larmor phase value ($\phi = \omega T = \frac{2\mu B}{\hbar}\frac{w}{v}$).

Applying the aforementioned analysis for a gaussian incident wavepacket, the final transmitted state is given by -  
\begin{equation}
    \begin{split}
        \ket{\Psi_{sc}} &= \psi(x,t) \otimes \frac{1}{\sqrt{2}}\left( \, \ket{\uparrow}_{z} + e^{i\phi} \, \ket{\downarrow}_{z}\right) \\ 
        \psi(x,t) &= \frac{1}{(2\pi\sigma^2)^{1/4}} e^{i(k_0x-\omega_0 t)} e^{-\frac{(x - x_0 - v_g t)^2}{4\sigma^2}}
    \end{split}
\end{equation}

From this resulting final state with $\braket{\psi^{+}}{\psi^{-}}=e^{i\phi}\int_{-\infty}^{\infty}\abs{\psi(x,t)}^2 dx = e^{i\phi}$, the value of coherence parameter is
\begin{equation}
    \mathcal{O} \equiv \expval{\sigma_x}^2 + \expval{\sigma_y}^2 = \abs{\braket{\psi^{+}}{\psi^{-}}}^2 = e^{-i\phi}e^{i\phi}=1 \,\,.
\end{equation}

\subsection{Spatial-spin mixed state} \label{app:mixed}
Decoherence or some unknown mechanism can produce a spatial-spin mixed state corresponding to the spin-embedded system emerging from the confined uniform magnetic field region. In general, such a mixed state can be written as 
\begin{equation} \label{eq:rho_mix}
\begin{aligned}
\rho_{\text{mix}}
&= p_{1} \, \ket{\uparrow_z}\bra{\uparrow_z} \otimes \ket{\psi^{+}}\bra{\psi^{+}} + 
p_{2} \, \ket{\uparrow_z}\bra{\uparrow_z} \otimes \ket{\psi^{-}}\bra{\psi^{-}} \\
&\quad + p_{3} \, \ket{\downarrow_z}\bra{\downarrow_z} \otimes \ket{\psi^{+}}\bra{\psi^{+}} + 
p_{4} \, \ket{\downarrow_z}\bra{\downarrow_z} \otimes \ket{\psi^{-}}\bra{\psi^{-}} \,\,,
\end{aligned}
\end{equation}
where, $p_1 + p_2 + p_3 + p_4 = 1$. Since we are performing only spin measurements, we trace out the spatial degree of freedom and calculate the probability of measuring spin along $+x$, \textit{i.e} $\ket{\chi} = \frac{1}{\sqrt{2}}\left( \ket{\uparrow_z} + \ket{\downarrow_z} \right)$:
\begin{equation} \label{eq:rho_mix_spin}
\rho_{\rm mix}^{\rm spin}
=\mathrm{Tr}_{\rm space}[\rho_{\text{mix}}]
= 
\begin{pmatrix}
p_1 + p_2 & 0 \\
0 & p_3 + p_4
\end{pmatrix}
\end{equation}
\begin{equation}
\Pi_{+x} = \ket{\chi}\bra{\chi}
= \frac{1}{2}
\begin{pmatrix}
1 & 1 \\
1 & 1
\end{pmatrix}
\end{equation}
Therefore,
\begin{equation}
\begin{aligned}
P(+x) &= \mathrm{Tr}\!\left( \rho_{\rm mix}^{\rm spin}\,\Pi_{+x} \right) \\
&= \mathrm{Tr}\!\left(
\begin{pmatrix}
p_1 + p_2 & 0 \\
0 & p_3 + p_4
\end{pmatrix}
\cdot
\frac{1}{2}
\begin{pmatrix}
1 & 1 \\
1 & 1
\end{pmatrix}
\right) \\
&= \frac{p_1 + p_2 + p_3 + p_4}{2} = \frac{1}{2} \,.
\end{aligned}
\end{equation}
This shows that regardless of the initial kinetic energy of the system a spatial-spin mixed state gives a probability of $\tfrac{1}{2}$ for measuring spin along $+x$. This holds true for any arbitrary mixture. \vspace{2mm}

Furthermore, using the reduced density matrix, $\rho_{\rm mix}^{\rm spin}$ given in Eq.~(\ref{eq:rho_mix_spin}), we calculate the following spin-expectation values,
\begin{equation}
\begin{aligned}
\expval{\sigma_x} &= \frac{1}{2} \left( 0+0 \right) = 0 \,\,, \\
\expval{\sigma_y} &= \frac{i}{2} \left( 0-0 \right) = 0 \,\,.
\end{aligned}
\end{equation}
Then, for an arbitrary mixed state, the value of the coherence parameter $\mathcal{O}$ turns out to be,
\begin{equation}
\mathcal{O} \equiv \expval{\sigma_x}^{2} + \expval{\sigma_y}^{2} = 0 \,\,.
\end{equation}

\newpage
\section{Alternate analysis using spin-probabilities} \label{app:probs_analysis}

\subsection{For Sharp Potential} \label{app:probs_analysis_sharp}
Using the state given in Eq.~\ref{eq:c10}, one can find the distribution of spins along $\ket{\chi} = \frac{1}{\sqrt{2}}\left(\ket{\uparrow}_{z} + \ket{\downarrow}_{z}\right)$ as
\begin{equation} \label{eq:c11}
P_{\chi}^{\prime} = \int \left| \braket{\chi}{\Psi_{q}} \right|^{2} \, dx = \int \rho_{\chi}^{\prime}(x) \, dx \,\, .
\end{equation}
Here, $\rho_{\chi}^{\prime}(x) = \left| \frac{1}{2} \left( \psi^{+}_{T}(x) + \psi^{-}_{T}(x) \right) \right|^{2}$, represents the local probability density for spin outcome $\ket{\chi}$. Now, the corresponding probability obtained by following the standard (semiclassical) treatment of Larmor precession, in this case, takes the form
\begin{equation} \label{eq:c12}
P_{\chi} = \int \left| \braket{\chi}{\Psi_{sc}} \right|^{2} \, dx = \int \rho_{\chi}(x) \, dx \,\, ,
\end{equation}
where, $\ket{\Psi_{sc}} = \psi_{0}(x) \, \frac{1}{\sqrt{2}} \left(\ket{\uparrow}_{z} + e^{i \phi} \ket{\downarrow}_{z}\right)$ (analogous to Eq.~\ref{eq:c6}), and $\psi_{0}(x) = \frac{1}{(2\pi\sigma^2)^{1/4}} e^{-(x - x_c)^2 / 4\sigma^2} e^{ik_0 x}$. Thus, $\rho_{\chi}(x) = \left| \frac{1}{2} \psi_{0}(x) \left( 1 + e^{i \phi} \right) \right|^{2}$. Note that $\phi = \frac{2 |\mu B_{0}| w}{\hbar v}$. \vspace{2mm}

We simplify and split up the integral in Eq.~\ref{eq:c11} into the following parts,
\begin{equation} \label{eq:P'_parts}
P_{\chi}^{\prime} = \frac{1}{4} \left( \text{IP} + \text{IP}^{*} + T_{+} + T_{-} \right) \,\,.
\end{equation}
Then, using Eq.~\ref{eq:P'_parts_solved_integrals}, we can express $P_{\chi}^{\prime}$ as
\begin{equation} \label{eq:P'_final_form}
\begin{aligned}
P_{\chi}^{\prime} &= \frac{|C_{+}|^{2}}{8} \,\, \textrm{erfc}\left(\frac{\frac{w}{2}-x_{c}}{\sqrt{2}\,\sigma}\right) + \frac{|C_{-}|^{2}}{8} \,\, \textrm{erfc}\left(\frac{\frac{w}{2}-(x_{c}+d)}{\sqrt{2}\,\sigma}\right) \\
&\quad + \frac{1}{4} \Re{C_{+}^{*} C_{-}} \, e^{-d^{2}/8\sigma^{2}} \, \textrm{erfc}\left(\frac{\frac{w}{2}-\left(x_{c}+\frac{d}{2}\right)}{\sqrt{2}\,\sigma}\right)
\end{aligned}
\end{equation}
Similarly, upon simplifying and solving the integral in Eq.~\ref{eq:c12}, we obtain the following expression for $P_{\chi}$,
\begin{equation} \label{eq:P_final_form}
P_{\chi} = \frac{1}{2} \cos^{2}\left(\frac{\phi}{2}\right) \, \textrm{erfc}\left(\frac{\frac{w}{2}-x_{c}}{\sqrt{2}\, \sigma}\right) \,\,.
\end{equation}
Using the $\textrm{erfc}(\,)$ simplification, we can rewrite the probabilities as
\begin{equation} \label{eq:simplified_probs}
\begin{aligned}
&P_{\chi}^{\prime} \approx \frac{|C_{+}|^{2}}{4} + \frac{|C_{-}|^{2}}{4} + \frac{1}{2} \Re{C_{+}^{*} C_{-}} \, e^{-d^{2}/8\sigma^{2}} \, , \\
&P_{\chi} \approx \cos^{2}\left(\frac{\phi}{2}\right) \, .
\end{aligned}
\end{equation}
Here, $|C_{\pm}|^{2}$ are as given in Eq.~\ref{eq:C+/-_squared}. Furthermore, using Eq.~\ref{eq:factor_C+&C-}, we obtain, simplify and rewrite $\Re{C_{+}^{*} C_{-}}$ as
\begin{equation} \label{eq:real(factor_plus_minus)}
\Re{C_{+}^{*} C_{-}} = \frac{16\,k^{2}k_{-}k_{+}}{\mathcal{D}_{1}\mathcal{D}_{2}}\big(\mathcal{N}_{1}+\mathcal{N}_{2}-\mathcal{N}_{3}+\mathcal{N}_{4}\big) \,\, ,
\end{equation}
such that,
\begin{equation}
\begin{aligned}
&\mathcal{D}_{1} \;=\; \big((k-k_{-})^{4}\sin^{2}(2wk_{-})+\big((k-k_{-})^{2}\cos(2wk_{-})-(k+k_{-})^{2}\big)^{2}\big), \\[6pt]
&\mathcal{D}_{2} \;=\; \big((k+k_{+})^{4}\sin^{2}(2wk_{+})+\big((k-k_{+})^{2}-(k+k_{+})^{2}\cos(2wk_{+})\big)^{2}\big),
\\[8pt]
&\mathcal{N}_{1} \;=\; (k-k_{-})^{2}(k+k_{+})^{2}\,\sin(2wk_{-})\,\sin(2wk_{+})\,\cos\!\big(wk_{-}+wk_{+}\big),
\\[6pt]
&\mathcal{N}_{2} \;=\; (k-k_{-})^{2}\Big((k-k_{+})^{2}-(k+k_{+})^{2}\cos(2wk_{+})\Big)\,\sin(2wk_{-})\,\sin\!\big(wk_{-}+wk_{+}\big),
\\[6pt]
&\mathcal{N}_{3} \;=\; (k+k_{+})^{2}\Big((k-k_{-})^{2}\cos(2wk_{-})-(k+k_{-})^{2}\Big)\,\sin(2wk_{+})\,\sin\!\big(wk_{-}+wk_{+}\big),
\\[6pt]
&\mathcal{N}_{4} \;=\; \Big((k-k_{-})^{2}\cos(2wk_{-})-(k+k_{-})^{2}\Big)\Big((k-k_{+})^{2}-(k+k_{+})^{2}\cos(2wk_{+})\Big)\,\cos\!\big(wk_{-}+wk_{+}\big).
\end{aligned}
\end{equation}
Additionally, for measurement of spin along some arbitrary $\ket{\theta} = \frac{1}{\sqrt{2}}\left(\ket{\uparrow}_{z}+e^{i\theta}\ket{\downarrow}_z\right)$, the corresponding probabilities would be of the following forms,
\begin{equation} \label{eq:probs_theta}
\begin{aligned}
&P_{\theta}^{\prime} \approx \frac{|C_{+}|^{2}}{4} + \frac{|C_{-}|^{2}}{4} + \frac{1}{2} \Re{e^{-i\theta} C_{+}^{*} C_{-}} \, e^{-d^{2}/8\sigma^{2}} \, , \\
&P_{\theta} \approx \cos^{2}\left(\frac{\phi-\theta}{2}\right) \, .
\end{aligned}
\end{equation}

\subsubsection{Convergence of the quantum and the semiclassical treatments} \label{app:gaussian_validity_sc&q}

We examine the high kinetic energy regime at which our full quantum treatment of the situation converges to the semiclassical treatment. Therefore, in this section we focus in the limit $K\gg |\mu B_{0}|$ \textit{i.e.} $r \ll 1$. In this limit, we will have $k_{+} \approx k_{-} \approx k$, and $d \to 0$. \vspace{1mm}

In tune with this, in Eq.~\ref{eq:P'_final_form}, we set $k_+ = k_- = k$ everywhere except inside $\sin(\,)$ and $\cos(\,)$ terms, as these are much more sensitive to variations in values of $k$, $k_+$, and $k_-$. By doing so, note that Eq.~\ref{eq:C+/-_squared} will reduce to: $|C_{+}|^{2} \to 1$, $|C_{-}|^{2} \to 1$. Similarly, Eq.~\ref{eq:real(factor_plus_minus)} will drastically simplify to $\Re{C_{+}^{*} C_{-}} \to \sin(2wk_{+}) \sin(wk_{+}+wk_{-}) + \cos(2wk_{+}) \cos(wk_{+}+wk_{-}) = \cos(wk_{-}-wk_{+})$.

Using the above results, Eq.~\ref{eq:P'_final_form} reduces to
\begin{equation} \label{eq:P'_highK_limit}
P_{\chi}^{\prime} \to \frac{1}{4} \left(1 + \cos(wk_{-}-wk_{+})\right) \, \textrm{erfc}\left(\frac{\frac{w}{2}-x_{c}}{\sqrt{2}\,\sigma}\right) = \frac{1}{2} \cos^{2}\left(\frac{wk_{-}-wk_{+}}{2}\right) \, \textrm{erfc}\left(\frac{\frac{w}{2}-x_{c}}{\sqrt{2}\,\sigma}\right) \,\,.
\end{equation}
Now, we binomially expand $k_+$ and $k_-$ around $k$ and keep terms to the order of $|\mu B_{0}|/K$ ignoring higher order terms. Upon doing this, we obtain $(k_{-}-k) = \frac{k}{2} \frac{|\mu B_{0}|}{K} = \frac{|\mu B_{0}|}{\hbar v}$, and $(k_{+}-k) = - \frac{|\mu B_{0}|}{\hbar v}$. Then, we rewrite Eq.~\ref{eq:P'_highK_limit} as
\begin{equation} \label{eq:P'_highK_limit_final}
P_{\chi}^{\prime} \to \frac{1}{2} \cos^{2}\left(\frac{|\mu B_{0}| w}{\hbar v}\right) \, \textrm{erfc}\left(\frac{\frac{w}{2}-x_{c}}{\sqrt{2}\,\sigma}\right) = \frac{1}{2} \cos^{2}\left(\frac{\phi}{2}\right) \, \textrm{erfc}\left(\frac{\frac{w}{2}-x_{c}}{\sqrt{2}\,\sigma}\right) = P_{\chi} \,\,.
\end{equation}
Invoking the $\textrm{erfc}(\,)$ simplification stipulated earlier in~\ref{app:gaussian_coh_para,etc}, we obtain (in the $r \ll 1$ regime),
\begin{equation} \label{eq:T+_T-_IPsq_highK_limit}
P_{\chi}^{\prime} \to P_{\chi} \approx \cos^{2}\left(\frac{\phi}{2}\right) \,,\,\,\, T_{+} \approx 1 \,,\,\,\, T_{-} \approx 1 \,\,,\text{and}\,\,\, \mathcal{O} \approx 1 \,\,.
\end{equation}
Therefore, in the high kinetic energy limit, the modified probability obtained from the full quantum (coupled) treatment of the problem reduces to the probability obtained in the semiclassical treatment ($P_{\chi}^{\prime} \to P_{\chi}$ when $r \ll 1$).

To understand these results physically, note that in the high kinetic energy regime ($r \ll 1$), the entire wave is transmitted but effectively picks up a phase. In tune with this, there is almost no spatial separation developed between the transmitted $\psi^{\pm}$ components as the time evolution of the entire wavefunction occurs due to a very shallow well and a very low barrier. Therefore, $\mathcal{O} \approx 1$ effectively follows from $d \to 0$. \vspace{2mm}

The following plot of the probabilities in the small-$r$ regime verifies the above findings numerically.

\begin{figure}[H]
    \centering
    \includegraphics[width=0.75\textwidth]{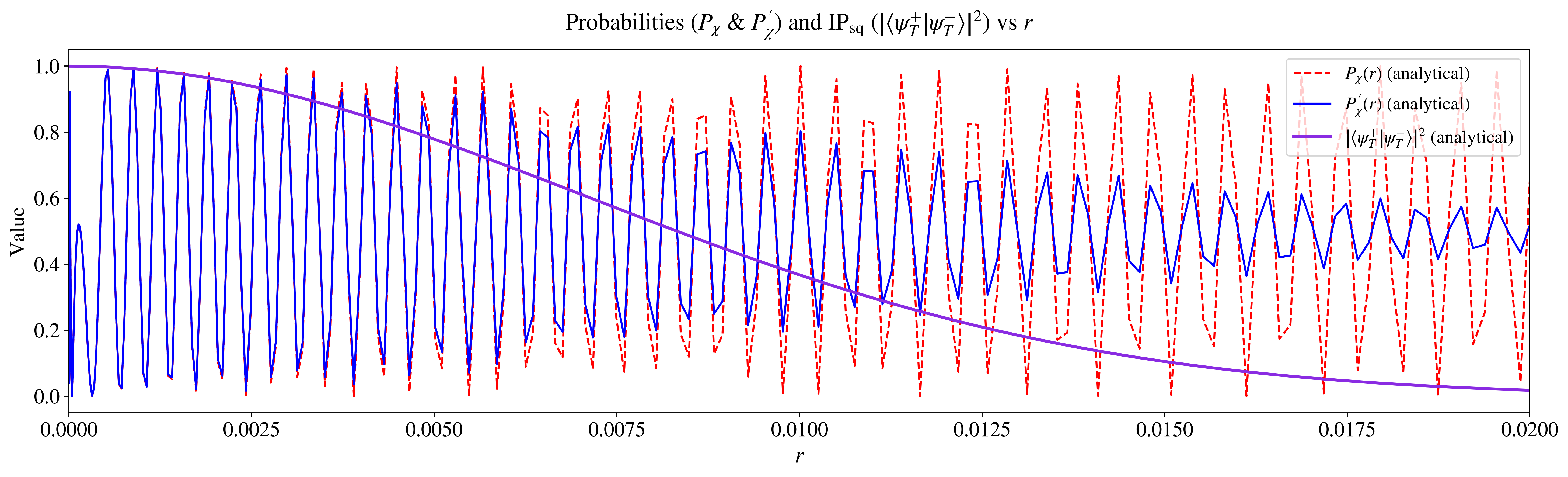}
    \caption{$P_{\chi}$, $P^{\prime}_{\chi}$, and $\mathcal{O}$ for small $r$ (Parameters: $m = 10^{3}$ amu, $w = 20~\mu\text{m}$, $\sigma = 100~\text{nm}$, and $B_{0} = 0.05~\text{T}$).}
    \label{fig:probs_small_r}
\end{figure}

\subsubsection{Effect of varying the initial wavepacket width on the dynamics} \label{app:data1_sigma}

In the scenario of the sharp potential, we further present the variation of $P^{\prime}$ and $\mathcal{O}$ vs $r$ (by varying the incident velocity) for different values of the wavepacket width $\sigma$, to illustrate how the spatial extent of the wavepacket influences these quantities. Before going into the plots, however, we present the following analysis. \vspace{1mm}

We reiterate here Eq.~\ref{eq:P'_parts}, which reads as
\begin{equation}
P_{\chi}^{\prime} = \frac{1}{4} \left( \text{IP} + \text{IP}^{*} + T_{+} + T_{-} \right) \,\,,
\end{equation}
The behavior of $P_{\chi}^{\prime}$ in the small-$r$ regime is already detailed in~\ref{app:gaussian_validity_sc&q}. Here, we describe the behavior in the large-$r$ regime. As explained earlier in~\ref{app:gaussian_coh_para,etc}, we note again that $x_{c} = \frac{w}{2} + \Delta x$, where $\Delta x \sim n \sigma$. This then simplifies the $\textrm{erfc}(\,)$ factor as $\textrm{erfc}\left(\frac{\frac{w}{2}-x_{c}}{\sqrt{2}\,\sigma}\right) = \textrm{erfc}\left(\frac{-n}{\sqrt{2}}\right) \approx 2$. Thus,
\begin{equation} \label{eq:P_moderate&high_r_limit}
P_{\chi} \approx \cos^{2}\left(\frac{\phi}{2}\right) \,\,.
\end{equation}
Importantly, note here that the definitions of "small" and "large" $r$ are evidently dependent on the scale of the state, \textit{\textit{i.e.}}, the spatial width of the Gaussian wavepacket, $\sigma$. Here, we emphasize that the trend of $P^{\prime}_{\chi}$ in the small $r$ regime is completely attributed to the inner product (overlap) terms, since $T_{\pm} \approx 1$ throughout that regime. Now, as $r$ is increased, $d$ increases and $\mathcal{O} \to 0$. Therefore, an objective way to identify the regime of large $r$ is by identifying the parametric space in which $\mathcal{O} \approx 0$. For Gaussian wavepackets, it is known that $\sim 99\%$ of the packet is contained within $x_{c} \pm 2.5 \, \sigma$; $x_{c}$ being the center of the packet. Then, for $\frac{d}{\sigma} > 5$, we will get $\mathcal{O} \approx 0$. Then, the large-$r$ regime is identified where,
\begin{equation} \label{eq:criterion_moderate&high_r}
\frac{w}{5\sigma} \left( \frac{1}{\sqrt{1-r}} - \frac{1}{\sqrt{1+r}} \right) = \frac{w}{5\sigma} \left( \sqrt{\frac{K}{K-|\mu B_0|}} - \sqrt{\frac{K}{K+|\mu B_0|}} \right) > 1
\end{equation}
Conversely, consistent with this definition, the small-$r$ regime is identified by $0 < \mathcal{O} \leq 1$; equivalently when $d < 5\sigma$. This criterion provides a parametric demarcation between the small-, and large-$r$ regimes (for example, if $\sigma=1~\mu\text{m}$ and $w=20~\mu\text{m}$, we may refer to $r \gtrapprox 0.25$ as the regime of large $r$). In this considered regime, Eq.~\ref{eq:criterion_moderate&high_r} holds, such that $e^{-d^{2}/8\sigma^{2}} \to 0$ and $e^{-d^{2}/4\sigma^{2}} \to 0$, thereby suppressing $\textrm{IP}$, $\textrm{IP}^{*}$, and $\mathcal{O}$.
\begin{equation} \label{eq:IP&IPsq_moderate_high_r}
\textrm{IP} \to 0 \,,\,\,\, \textrm{IP}^{*} \to 0 \,,\, \text{and} \,\, \mathcal{O} \to 0 \,.
\end{equation}
Using these simplifications, we rewrite $P_{\chi}^{\prime}$ as
\begin{equation} \label{eq:P'_moderate&high_r_limit}
P_{\chi}^{\prime} \to \frac{1}{4} \left(T_{+}+T_{-}\right)
\end{equation}
Within the $0.25 \lessapprox r \lessapprox 0.6$ sub-regime, we will have good transmittance of both $\psi^{\pm}$ components. Therefore within this regime, we may approximate $T_{\pm} \approx 1$. Thus,
\begin{equation} \label{eq:P'_moderate_r_limit}
P_{\chi}^{\prime} \approx \frac{1}{2} \,\,.
\end{equation}
This behavior manifests in the numerical results as well, as signified by the saturation in the values of $P_{\chi}^{\prime}$ around $0.5$.

In the $r \gtrapprox 0.6$ sub-regime, as $r \to 1$, the $\psi^{+}$ component gradually becomes slower and slower, and $T_{+} \to 0$ in an oscillatory manner. $T_{-}$ too falls but only marginally as compared to $T_{+}$. As a result of this, from Eq.~\ref{eq:P'_moderate&high_r_limit}, we have $P_{\chi}^{\prime}$ falling off oscillatorily to around a value of 0.25. Thus,
\begin{equation} \label{eq:P'_high_r_limit}
 \frac{1}{4} \lessapprox P_{\chi}^{\prime} \lessapprox \frac{1}{2} \quad \text{(as $r\to 1$)} \,\,.
\end{equation}
\vspace{2mm}In light of these understandings, we now present the following plots:

\begin{figure}[H]
    \centering
    \includegraphics[width=\textwidth]{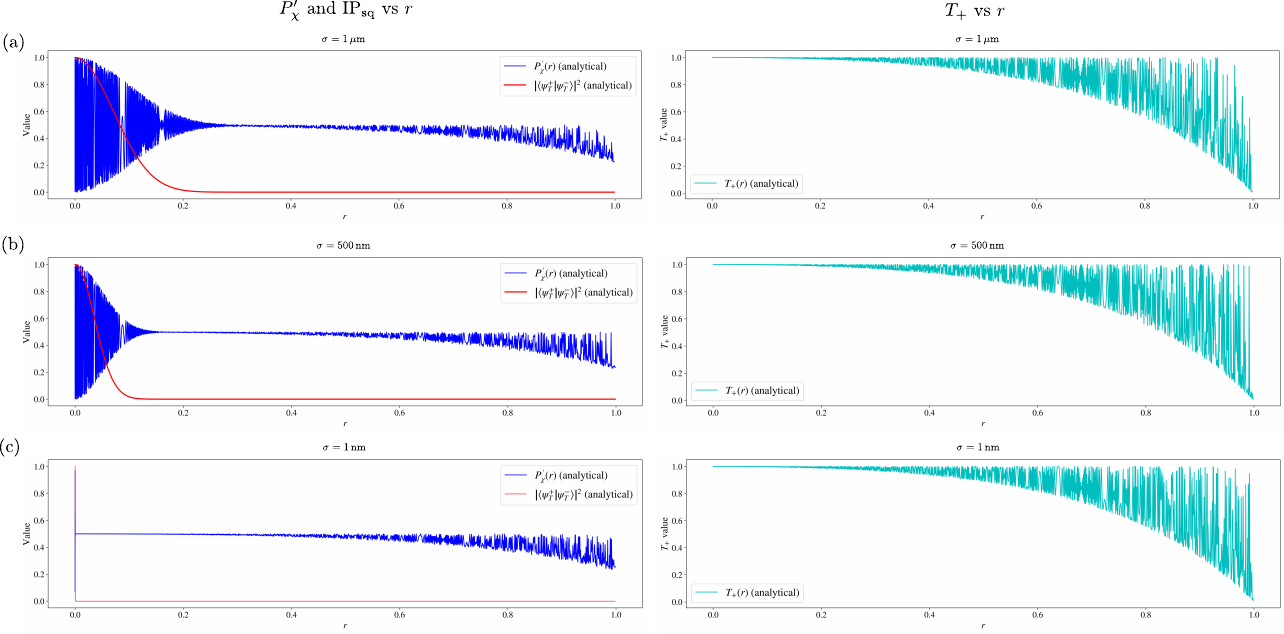}
    \caption{Variaton of $P^{\prime}$, $\mathcal{O}$, and $T_{+} = \int |\psi_{T}^{+}|^{2} \, dx$ vs $r$ for (a) $\sigma = 1 \,\mu\text{m}$, (b) $\sigma = 500 \,\text{nm}$, and (c) $\sigma = 1 \,\text{nm}$. Other relevant parameters are fixed at $m = 10^{6}$ amu, $w = 20\, \mu\text{m}$, and $B_{0} = 0.05$ T.}
    \label{fig:diff_sigma}
\end{figure}
\noindent It is observed that the oscillations in $P^{\prime}$ persist within the regime where $\mathcal{O} > 0$, as expected. As $\mathcal{O} \rightarrow 0$, these oscillations gradually diminish and stabilize around $P^{\prime} \approx 0.5$. Oscillatory behavior reemerges only for larger values of $r > 0.5$, which is a direct consequence of the transmission coefficient of the $\psi^{+}$ component ($T_{+}$) exhibiting an oscillatory decay towards zero with increasing $r$.

\subsection{For Smooth Potential} \label{app:probs_analysis_smooth}

The smooth potential scenario carries an additional nuance when $r$ is close to 1 ($K \sim \mu B_0$). The solutions in~\ref{app:approx wkb} are derived in the dominant-transmission regime and neglect above-barrier reflection. Therefore, we provide a justification of this approach for large masses, and derive an analytical upper bound on the above-barrier reflection probability. The double-tanh barrier profile possesses two well-separated edges for all parameters considered in this work ($sw \geq 2$), each of which takes the form of a smooth potential step,
\begin{equation}
    V_\mathrm{edge}(\xi) = \frac{\mu B_0}{1 + e^{-2s\xi}} = \frac{\mu B_0}{2}(1+\tanh(s \xi)),
\end{equation}
where $\xi$ measures displacement from the respective edge. The exact reflection coefficient for this smooth step potential is given by Landau and Lifshitz (Vol. 3, §25, Problem 3) \cite{LandauLifshitzQM} as
\begin{equation} \label{eq:eckart_exact}
    |R_\mathrm{edge}|^2 = 
    \frac{\sinh^2\!\left(\dfrac{\pi k_0(1 - \sqrt{1-r})}{2s}\right)}
         {\sinh^2\!\left(\dfrac{\pi k_0(1 + \sqrt{1-r})}{2s}\right)},
\end{equation}
where $k_0 = \sqrt{2m\mu B_0 / r} \, / \hbar$ is the asymptotic wavenumber and $r = \mu B_0 / K$. Due to symmetry, the rising (left) and falling (right) edges yield identical $|R_\mathrm{edge}|^2$. Therefore, the total reflected amplitude is bounded by,
\begin{equation}
    |R_\mathrm{total}|^2 \leq 4\,|R_\mathrm{edge}|^2,
    \label{eq:R_bound}
\end{equation}
where the factor of four represents worst-case scenario of constructive interference between the two edges.\vspace{1mm}

\textit{Smooth-edge regime with dominant transmission ($k_0/s \gg 1$)}: When the de~Broglie wavelength is much shorter than the edge width, both sinh arguments in Eq.~\eqref{eq:eckart_exact} are large ($\sinh x \approx e^x/2$), and the reflection probability reduces to
\begin{equation}
    |R_\mathrm{edge}|^2 \approx 
    \exp\!\left(-\frac{2\pi k_0 \sqrt{1-r}}{s}\right).
    \label{eq:R_smooth}
\end{equation}

The key dimensionless parameter governing above-barrier reflection 
is $k_0/s$, the ratio of the asymptotic wavenumber to the barrier 
edge steepness. Evaluating this ratio explicitly,
\begin{equation}
    \frac{k_0}{s} = \frac{1}{\hbar s}\sqrt{\frac{2m\mu B_0}{r}},
    \label{eq:k0s}
\end{equation}
The condition $k_0/s \gg 1$ is satisfied for systems such as NV-nanodiamonds (which is the system of primary interest in the main text), with $\mu \gtrsim \mu_B$ ($\mu_B$ is Bohr magneton) for any mass $m \geq 100$ amu, throughout $s \in (0.1,\,1)\,\mu\mathrm{m}^{-1}$ when the magnetic field strength is as low as $5 \mu T$. These parameter values place the system well within the smooth-edge regime where the reflection exponent satisfies $\Gamma \equiv 2\pi k_0\sqrt{1-r}/s \gg 1$ throughout the "small-r" and "large-r" parameter ranges studied in this work (excluding when $r \approx 1$), giving $|R_\mathrm{total}|^2 \ll 1$ and justifying the dominant transmission WKB solution. For the specific system studied in this work ($m = 10^{10}$\,amu, $\mu = \mu_\mathrm{NV}, s = 0.1 ~\mu m^{-1}, B_0 = 5 ~\mu T$), $k_0/s \sim 10^6$ and $\Gamma \gtrsim 10^6$ even at $r = 0.99$, the reflection is therefore entirely negligible ($|R_\mathrm{total}|^2 < 4e^{-10^6}$). \vspace{2mm}

If we consider a system whose magnetic moment is such that $\mu \gtrsim \mu_N$ ($\mu_N$ is nuclear magneton) then reflection is negligible throughout most of the $r \lesssim 0.95$ regime for $m \gtrsim 10^4$ amu. But, for lighter particles or smaller magnetic moments, $k_0/s$ can fall below one. Here, the smooth-edge approximation breaks down and above-barrier reflection grows substantially as $r \to 1$. This scenario can be solved by simulating the TDSE with the smooth-edged field using $m = 10^3$ amu and $\mu = -1.913 \mu_N$. As a consistency check, we first recover the sharp potential results by taking the large-$s$ limit ($s= 20 ~\mu m^{-1}$) and computing the observables $|\Delta P_{\textrm{c}}|$, $d$, and $\mathcal{O}$ (Table \ref{tab:num_s20_r}). The spatial domain is discretized on a uniform grid, and temporal evolution is carried out using the Crank-Nicolson method. We then analyze the genuinely smooth case by setting the steepness parameter to $s = 1\, (\mu\text{m})^{-1}$ and observe the fall off of $P_\chi$ due to reflection at $r \rightarrow 1$ (Table \ref{tab:num_s1_r}). While this method is tractable up to masses of $10^5$ amu, it becomes computationally expensive at higher masses.

\subsection{Key features} \label{app:key_features_probs}

\begin{figure*}[h!]
    \centering
    \includegraphics[width=\textwidth]{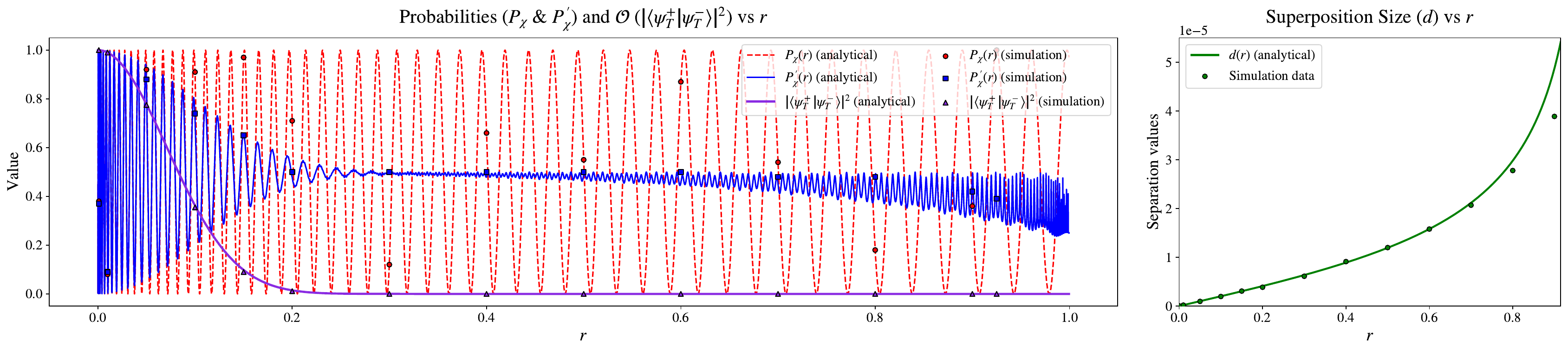}
    \caption{Illustrative data for $P_{\chi}$, $P^{\prime}_{\chi}$, $\mathcal{O}$, and $d$ versus $r$ for a sharp potential. Curves show analytical results, while discrete markers denote sample simulation data (for $s = 20\,(\mu\mathrm{m})^{-1}$). Parameters: $m = 10^{3}$ amu, $w = 20\,\mu\mathrm{m}$, $\sigma = 1\,\mu\mathrm{m}$, $B_{0} = 10^{-4}$ T, $\mu = -\mu_{\rm n} = 1.913 \times 5.050783699\times 10^{-27}~\text{J\,.\,T}^{-1}$.
    }
    \label{fig:plots_with_data}
\end{figure*}

We first note the key observations from our results (for detailed results see~\ref{app:data}): 
\begin{enumerate}
    \item In the small $r$ (high kinetic energy, $K$) regime, $P^{\prime}_{\chi}$ converges to $P_{\chi}$ in an oscillatory manner such that $\lvert \Delta P_{\textrm{c}} \rvert \to 0$ as $r \to 0$ (see Apps.~\ref{app:gaussian_validity_sc&q},~\ref{app:approx gaussian}).
    \item In the large $r$ regime, $P^{\prime}_{\chi}$ first saturates to around a value of 0.5, as $\mathcal{O} \to 0$. Subsequently, $P^{\prime}_{\chi}$ starts to fall off and oscillates between 0.5 and 0.25, as the transmission for the spin-up component starts to fall off oscillatorily (see~\ref{app:data1_sigma}). These observations are fundamentally distinct from the semiclassical treatment, where $P_{\chi}$ continues to oscillate.
    \item We note that the fall off in $P^{\prime}_{\chi}$ as $r \to 1$ is much suppressed when passing through a smooth potential.
    \item In the large $r$ regime, we can generate $d \gg \sigma$.  This feature from our treatment is of genuine interest since such a spatial separation is also \textit{macroscopic} in terms of the superposition size, beyond fundamentally involving \textit{massive} systems. We emphasize that no such superposition arises in the semiclassical treatment, where spatial-spin coupling is absent.
    \item Furthermore, the above features persist for higher masses as well. In particular, both the deviation between $P^{\prime}_{\chi}$ and $P_{\chi}$ in the relevant parameter regimes and the emergence of a finite spatial separation $d$ remain robust, demonstrating that our scheme is, in principle, scalable to larger masses.
\end{enumerate}

\subsection{Data} \label{app:data}
Key parameters:
$r \equiv |\mu B_{0}|/K$, $t_{f}$ is the total time of evolution of wavefunction, $P' = \left| \braket{\chi}{\Psi_{q}} \right|^{2}$, $P = \left| \braket{\chi}{\Psi_{sc}} \right|^{2}$, $\lvert \Delta P_{\textrm{c}} \rvert = \lvert(P_{\chi}^{\prime}-P_{\chi})\rvert$, $d$ = spatial separation between Gaussian peaks of $\psi^{\pm}$, $\mathcal{O} = \left|\braket{\psi^{+}}{\psi^{-}}\right|^{2}$, and the values of natural constants used is as follows: $\hbar = 1.054571817 \times 10^{-34}~\text{J\,.\,s}, \, \text{1 amu} = 1.6605390666\times 10^{-27}~\text{Kg}, \, \mu_{\rm n} = 1.913 \times 5.050783699\times 10^{-27}~\text{J\,.\,T}^{-1}, \, k_{\mathrm{B}} = 1.380649 \times 10^{-23}~\text{J\,.\,K}^{-1}$

\subsubsection{Evidencing MQC by varying incident velocity} \label{app:data1}

We vary the initial velocity of the system, while keeping all other parameter values fixed (for instance: $m=10^{3}$ amu, $B_{0} = 10^{-4}$ T, $w=20 \, \mu m$, $s = 20\,(\mu m)^{-1}$, $\sigma=1\,\mu m$). This amounts to varying the initial kinetic energy $K$ which in turn amounts to varying $r$ (since $\mu B_0$ is fixed). From the data [\ref{app:data1}, Tables~\ref{tab:anlyt_r},\ref{tab:num_s20_r}], we observe that in the high velocity regime ($r\ll0.1$), where the kinetic energy dominates over the spin magnetic field potential energy, $P_{\chi}^{\prime} \approx P_{\chi}$ and $\lvert \Delta P_{\textrm{c}} \rvert\approx0$ (detailed analysis of this regime is presented in~\ref{app:gaussian_validity_sc&q}). Also, the value of $\mathcal{O}$ is around 1, and $d$ remains smaller than or comparable to $\sigma$. But as the initial velocity of the fired particles is decreased ($0.1<r$ and $r \rightarrow 1$) the kinetic energy of the particle becomes comparable to its potential energy and in this case we observe that $\lvert \Delta P_{\textrm{c}} \rvert \neq 0$ because $P_{\chi}^{\prime}$ oscillatorily saturates to value close to 0.5 while the value of $P_{\chi}$ keeps oscillating (see Fig.~\ref{fig:plots_with_data}). Thus, we not only observe a deviation but a completely different trend because of the presence of spatial-spin coupling. We also note that as $r \rightarrow 1$ the value of $\mathcal{O}$ smoothly falls to 0 and the value of $d$ smoothly rises to around $d \sim 10^{-5}$--$10^{-4}$ m. This can be explained by the fact that since one component experiences a potential barrier while the other experiences a potential well so while passing through the magnetic field region both components have different wavenumbers (and different speeds) and one component ($\psi^{-}$) traverses further than the other ($\psi^{+}$) and after exiting the magnetic field region both components evolve freely thus the separation between the two components persists.

\begin{table}[h]
\centering
\setlength{\tabcolsep}{9pt}
\renewcommand{\arraystretch}{1.0}
\begin{tabular}{cccccccc}
\toprule
$r$ & Temp (K) & $t_{f}$ (s) & $P'_{\chi}$ & $P_{\chi}$ & $\lvert \Delta P_{\textrm{c}} \rvert$ & $d$ (m) & $\mathcal{O}$ \\
\midrule
\num{0.001} & \num{4.66e-05} & 0.0009 & 0.38 & 0.38 & 0.00 & \num{2.00e-08} & \num{1.00e+00} \\
\num{0.01}  & \num{4.66e-06} & 0.0028 & 0.08 & 0.08 & 0.00 & \num{2.00e-07} & \num{9.90e-01} \\
\num{0.05}  & \num{9.33e-07} & 0.0063 & 0.86 & 0.92 & 0.06 & \num{1.00e-06} & \num{7.77e-01} \\
\num{0.1}   & \num{4.66e-07} & 0.0091 & 0.72 & 0.91 & 0.19 & \num{2.01e-06} & \num{3.62e-01} \\
\num{0.2}   & \num{2.33e-07} & 0.0134 & 0.48 & 0.71 & 0.23 & \num{4.10e-06} & \num{1.47e-02} \\
\num{0.3}   & \num{1.56e-07} & 0.0172 & 0.50 & 0.12 & 0.38 & \num{6.36e-06} & \num{3.97e-05} \\
\num{0.4}   & \num{1.17e-07} & 0.0210 & 0.48 & 0.66 & 0.18 & \num{8.92e-06} & $\sim$ \num{0} \\
\num{0.5}   & \num{9.33e-08} & 0.0251 & 0.49 & 0.55 & 0.06 & \num{1.20e-05} & $\sim$ \num{0} \\
\num{0.6}   & \num{7.78e-08} & 0.0299 & 0.46 & 0.87 & 0.41 & \num{1.58e-05} & $\sim$ \num{0} \\
\num{0.7}   & \num{6.66e-08} & 0.0361 & 0.45 & 0.54 & 0.09 & \num{2.12e-05} & $\sim$ \num{0} \\
\num{0.8}   & \num{5.83e-08} & 0.0454 & 0.48 & 0.18 & 0.30 & \num{2.98e-05} & $\sim$ \num{0} \\
\num{0.9}   & \num{5.18e-08} & 0.0644 & 0.49 & 0.36 & 0.13 & \num{4.87e-05} & $\sim$ \num{0} \\
\num{0.925}   & \num{5.04e-08} & 0.0740 & 0.37 & 1.00 & 0.63 & \num{5.86e-05} & $\sim$ \num{0} \\
\num{0.975}   & \num{4.78e-08} & 0.1249 & 0.28 & 0.12 & 0.16 & \num{1.12e-04} & $\sim$ \num{0} \\
\bottomrule
\end{tabular}%

\caption{Analytical data: $m=10^{3}$ amu, sharp potential, $w = 20~\mu m$, $B_0 = 10^{-4}$ T, $\mu = -\mu_{\rm n}$}
\label{tab:anlyt_r}
\end{table}

\begin{table}[h]
\centering
\setlength{\tabcolsep}{9pt}
\renewcommand{\arraystretch}{1.0}
\begin{tabular}{cccccccc}
\toprule
$r$ & Temp (K) & $t_{f}$(s) & $P'_{\chi}$ & $P_{\chi}$ & $|\Delta P_{\textrm{c}}|$ & $d$ (m) & $\mathcal{O}$ \\
\midrule
\num{0.001} & \num{4.67e-05} & 0.002 & 0.37 & 0.38 & 0.01 & \num{2.00e-08} & \num{1.00e+00} \\
\num{0.01}  & \num{4.67e-06} & 0.003 & 0.09 & 0.08 & 0.01 & \num{1.99e-07} & \num{9.90e-01} \\
\num{0.05}  & \num{9.34e-07} & 0.006 & 0.88 & 0.92 & 0.04 & \num{9.67e-07} & \num{7.75e-01} \\
\num{0.1}   & \num{4.67e-07} & 0.016 & 0.74 & 0.91 & 0.17 & \num{1.97e-06} & \num{3.55e-01} \\
\num{0.15}  & \num{3.11E-07} & 0.024 & 0.65 & 0.97 & 0.32 & \num{3.08E-06} & \num{8.96E-02}   \\
\num{0.2}   & \num{2.34e-07} & 0.028 & 0.50 & 0.71 & 0.21 & \num{3.87e-06} & \num{1.12e-02} \\
\num{0.3}   & \num{1.56e-07} & 0.019 & 0.50 & 0.12 & 0.38 & \num{6.12e-06} & \num{1.28e-05} \\
\num{0.4}   & \num{1.17e-07} & 0.024 & 0.50 & 0.66 & 0.16 & \num{9.13e-06} & $\sim$ \num{0} \\
\num{0.5}   & \num{9.33e-08} & 0.033 & 0.50 & 0.55 & 0.05 & \num{1.20e-05} & $\sim$ \num{0} \\
\num{0.6}   & \num{7.78e-08} & 0.036 & 0.50 & 0.87 & 0.37 & \num{1.58e-05} & $\sim$ \num{0} \\
\num{0.7}   & \num{6.67e-08} & 0.042 & 0.48 & 0.54 & 0.06 & \num{2.07e-05} & $\sim$ \num{0} \\
\num{0.8}   & \num{5.84e-08} & 0.063 & 0.48 & 0.18 & 0.30 & \num{2.78e-05} & $\sim$ \num{0} \\
\num{0.9}   & \num{5.19e-08} & 0.118 & 0.42 & 0.36 & 0.06 & \num{3.89e-05} & $\sim$ \num{0} \\
\num{0.925} & \num{5.05E-08} & 0.118 & 0.39 & 1.00 & 0.61 & \num{4.28E-05} & $\sim$ \num{0}  \\
\bottomrule
\end{tabular}%

\caption{Simulation data: $m=10^{3}$ amu, $s$ = 20 $(\mu m)^{-1}$ (sharp potential), $w$ = 20 $\mu m$, $B_0 = 10^{-4}$ T, $\mu = -\mu_{\rm n}$}
\label{tab:num_s20_r}
\end{table}

\newpage
We find that much like our previous results of sharp barrier, the smooth barrier case shows the same trends in data. At high values of kinetic energy ($r\ll0.1$) $\lvert \Delta P_{\textrm{c}} \rvert \approx 0$ and as the kinetic energy is decreased ($r \rightarrow 1$) $P_{\chi}^{\prime}$ again saturates to a value close to 0.5 and at such $r$ values ($r>0.1$) we also have $\lvert \Delta P_{\textrm{c}} \rvert \neq 0$. Additionally similar to previous results $\mathcal{O}$ falls to 0 and $d$ increases as $r$ is increased. Thus, the deviations are persistent even in the cases where the boundary of magnetic field region is \textit{not} sharp. For all the results, the temperature values for the system are within the range of $10^{-8}$--$10^{-6}$ K, and the total evolution time ($t_{f}$) ranges from 0.011s to 0.118s. [\ref{app:data1}, Table~\ref{tab:num_s1_r}] \\

\begin{table}[h]
\centering
\setlength{\tabcolsep}{9pt}
\renewcommand{\arraystretch}{1.0}
\begin{tabular}{cccccccc}
\toprule
$r$ & Temp (K) & $t_{f}$(s) & $P'_{\chi}$ & $P_{\chi}$ & $|\Delta P_{\textrm{c}}|$ & $d$ (m) & $\lvert \langle \psi^+ | \psi^- \rangle \rvert^2$ \\
\midrule
0.01 & \num{4.67e-06} & 0.011 & 0.10 & 0.08 & 0.02 & \num{1.95e-07} & \num{9.88e-01} \\
0.1  & \num{4.67e-07} & 0.017 & 0.75 & 0.91 & 0.16 & \num{2.00e-06} & \num{3.43e-01} \\
0.2  & \num{2.34E-07} & 0.024 & 0.52 & 0.71 & 0.19 & \num{4.11E-06} & \num{9.90E-03}   \\
0.3  & \num{1.56E-07} & 0.047 & 0.50 & 0.12 & 0.38 & \num{6.22E-06} & $\sim 0$   \\
0.4  & \num{1.17E-07} & 0.050 & 0.50 & 0.66 & 0.16 & \num{8.59E-06} & $\sim 0$   \\
0.5  & \num{9.33E-08} & 0.053 & 0.50 & 0.55 & 0.05 & \num{1.17E-05} & $\sim 0$   \\
0.6  & \num{7.78E-08} & 0.055 & 0.50 & 0.87 & 0.27 & \num{1.52E-05} & $\sim 0$   \\
0.7  & \num{6.67E-08} & 0.057 & 0.50 & 0.54 & 0.04 & \num{2.03E-05} & $\sim 0$   \\
0.8  & \num{5.84E-08} & 0.063 & 0.49 & 0.18 & 0.31 & \num{2.62E-05} & $\sim 0$   \\
0.9  & \num{5.19E-08} & 0.118 & 0.48 & 0.36 & 0.12 & \num{3.53E-05} & $\sim 0$   \\
0.925 & \num{4.92E-08} & 0.118 & 0.46 & 1.00 & 0.54 & \num{3.85E-05} & $\sim 0$  \\
\bottomrule
\end{tabular}%

\caption{Simulation data: $m=10^{3}$ amu, $s$ = 1 $(\mu m)^{-1}$ (smooth potential), $w$ = 20 $\mu m$, $B_0 = 10^{-4}$ T, $\mu = -\mu_{\rm n}$}
\label{tab:num_s1_r}
\end{table}


\newpage
\section{Applying our framework to spin-embedded NV nanodiamonds} \label{app:NV}

\subsection{Analysis} \label{app:analysis_NV}
For a nanodiamond system with a single embedded spin in the NV-center, the interaction of magnetic field with electronic spin (spin-1 system) dominates over its interaction with nuclear spin. This is due to the higher magnetic moment of electronic spin as compared to the NV nuclear spin and thus the electronic spin plays a dominant role in the spin spatial coupling effects that our paper focuses on. The relevant Hamiltonian for our setup is as follows \cite{Pedernales2020}-
\begin{equation}
H = \frac{\hat{\mathbf P}^2}{2M} + \hbar D  S_z^2 -
\frac{\chi_m M}{2\mu_0}\,{\mathbf B}^2 + 
g_e \mu_B \,{\mathbf S}\cdot{\mathbf B}
\end{equation}
The above Hamiltonian consists of the kinetic term of COM motion (first term), the 2nd term accounts for the zero field splitting (ZFS) energy present in spin-1 systems, the third term refers to the diamagnetic interaction which encodes the repulsion experienced by the diamond (diamagnetic) in the presence of an external magnetic field and the last term corresponds to the electronic spin-field interaction, where $\mathbf{S}$ is the dimensionless spin operator.

Now if the initial electronic spin state is prepared in the state $\frac{1}{\sqrt{2}}(\ket{+1}+\ket{-1})$ then owing to the unitary dynamics we can reduce our Hamiltonian to the $\ket{\pm1}$ subspace. Thus, we essentially map from a spin-1 space to a spin-1/2 space, and it will still reflect the complete spatial-spin dynamics. Furthermore, we assume that the NV axis is aligned along the $z$ direction to avoid spurious spin rotations caused by misalignment with the magnetic field.
\begin{equation}
\Pi = \ket{+1}\bra{+1} + \ket{-1}\bra{-1} \, .
\end{equation}
\begin{equation}
\Pi \hat S_z^2 \Pi = \mathbb I_{2\times2} \,, \quad \& \quad \Pi \hat S_z \Pi =
\begin{pmatrix}
1 & 0 \\
0 & -1
\end{pmatrix}
\equiv \sigma_z \,.
\end{equation}
and the reduced Hamiltonian becomes -
\begin{equation}
H_{\mathrm{eff}}
=
\frac{\hat P_x^2}{2M}
+
\hbar D\,\mathbb I
-
\frac{\chi_m M}{2\mu_0}\,B^{2}(x)
-
\mu_{\rm NV} B(x)\,\sigma_z
\end{equation}
Now since the ZFS term essentially acts as a constant and contributes to a global energy phase shift. Therefore, we can rewrite the effective Hamiltonian as 
\begin{equation}
H_{\mathrm{eff}} = \frac{\hat P_x^2}{2M} - \frac{\chi_m M}{2\mu_0}\,B^{2}(x) - \mu_{\rm NV} B(x)\,\sigma_z
\end{equation}
Now, for simplicity, we shall take
\begin{equation}
    \frac{|\chi_m| M}{2\mu_0}\,{\mathbf B}^2 \ll |\mu_{\rm NV}| \mathbf{B} \implies B_0\ll\frac{2\mu_0|\mu_{\rm NV}|}{|\chi_m|M },
\end{equation}
Then, by ignoring diamagnetic effects in our present treatment, we have -
\begin{equation}
H_{\rm eff} = \frac{\hat{\mathbf P}^2}{2M} - \mu_{\rm NV} B(x)\,\sigma_z
\end{equation}
This effective Hamiltonian is similar to the one used in our treatment (ref.~Eq.~\ref{eq:Pauli_eqs & Hamiltonian}). Thus, our treatment and its solutions can be applied by changing to the magnetic moment of NV electronic spin ($\mu_{\rm NV} = 1.8548020157\times 10^{-23}~\text{J\,.\,T}^{-1}$ ). The caveat to this approach is that we impose a constraint on the maximum possible field strength to which our solutions apply for a given mass. For example, for a mass $m = 10^{7}$ amu, the maximum permissible field strength under which spin-field interaction dominates by two orders of magnitude would then be $B_0 \lesssim 5~\text{mT}$  allowing diamagnetic effects to be negligible at leading order. Also, in this case, we consider the wavepacket width of the nanodiamond system as $\sigma = 10~\text{nm}$ and a realisable coherence time of the embedded spin ($\tau \approx 1~\text{ms}$)~\cite{NV_coherence_time18,NV_coherence_time22,NV_coherence_time23}. \vspace{2mm}

\textit{Preliminary study of diamagnetism }-- Before presenting the results, we would also like to note the qualitative effect of inclusion of diamagnetic effect in our analysis. Since the diamagnetic interaction term is a spin-independent and $\textbf{B}^2$-dependent term, thus the effect of inclusion of this term would result in the shifting of the well and barrier height by an amount of $\mathcal{D}=\frac{|\chi_m| M}{2\mu_0}\,{\mathbf B}^2$ making the potential asymmetric. Including this term, the effective Hamiltonian for the two spin components reads -
\begin{equation}\label{eq:diamagH}
    H_{\mathrm{eff}} = \frac{\hat P_z^2}{2M} - \mu_{\rm NV} B(x)\,\sigma_z - \frac{\chi_m M}{2\mu_0}\,{\mathbf B}^2
\end{equation}
Consequently, accommodating the diamagnetic effect in the Hamiltonian modifies the separation $d$ of Eq.~\ref{eq:c16} in the sharp case as follows,
\begin{equation}\label{eq:d_diamag}
    d_{\rm mod} =w\left( \sqrt{\frac{K}{K - \mathcal{Z} - \mathcal{D}}} - \sqrt{\frac{K}{K + \mathcal{Z} - \mathcal{D}}} \right) =  w \left(\dfrac{1}{\sqrt{1 - r(\eta+1)}} - \dfrac{1}{\sqrt{1 - r(\eta - 1)}} \right) 
\end{equation}
where $\mathcal{Z} = |\mu_{\rm NV} B_0|$, $\eta = \mathcal{D}/\mathcal{Z}$ and $r = \mathcal{Z}/K$. Further, the transmission condition requires $K > \mathcal{D} + \mathcal{Z}$.  \vspace{1mm}

A key observation is that the difference between the potential experienced by spin-up and spin-down components, $V_+ - V_- = 2\mathcal{Z}$, is entirely independent of $\mathcal{D}$, since the diamagnetic term contributes identically to both potentials. Consequently, the spin-dependent splitting mechanism is never suppressed by $\mathcal{D}$. Instead, the diamagnetic term enhances $d$ within the transmission regime, rather than suppressing it. Both $\psi^{\pm}$ slow down equally inside the field, increasing the time the system spends in the confined field region and thus increasing the accumulated separation. The quantitatively relevant constraint affected by inclusion of diamagnetic interaction term $\mathcal{D}$ is not on the magnitude of $d$ but on the accessible r-parameter space bounded by the condition $K > \mathcal{D} + \mathcal{Z}$ or $r < 1/(\eta +1)$.



\begin{table}[H]
\centering
\setlength{\tabcolsep}{10pt}
\renewcommand{\arraystretch}{1.0}
\begin{tabular}{ccc}
\toprule
$r$ & $d$ & $d_{\rm mod}~(\eta=11.04)$ \\
\midrule
0.001 &   100  &   102 \\
0.005 &   500  &   544 \\
0.010 &  1000  &  1192 \\
0.020 &  2001  &  2909 \\
0.040 &  4006  &  9617 \\
0.060 &  6019  &  31207 \\
0.070 &  7022  &  68700 \\
\bottomrule
\end{tabular}
\caption{Comparison of the superposition size ($d_{\rm mod}$, $d$, given in nm) versus $r$ at $B_0 = 5~\mathrm{mT}$. $d_{\rm mod}$ incorporates the correction due to diamagnetic interaction ($\eta \approx 11.04$), while $d$ neglects the diamagnetic interaction. Parameters: $w = 100~\mu\mathrm{m}$, $M = 10^{10}~\mathrm{amu}$.}
\label{tab:dmod_sigma}
\end{table}

\begin{figure}[h]
\vspace{-1.5mm}
\centering
\includegraphics[width=0.6\columnwidth]{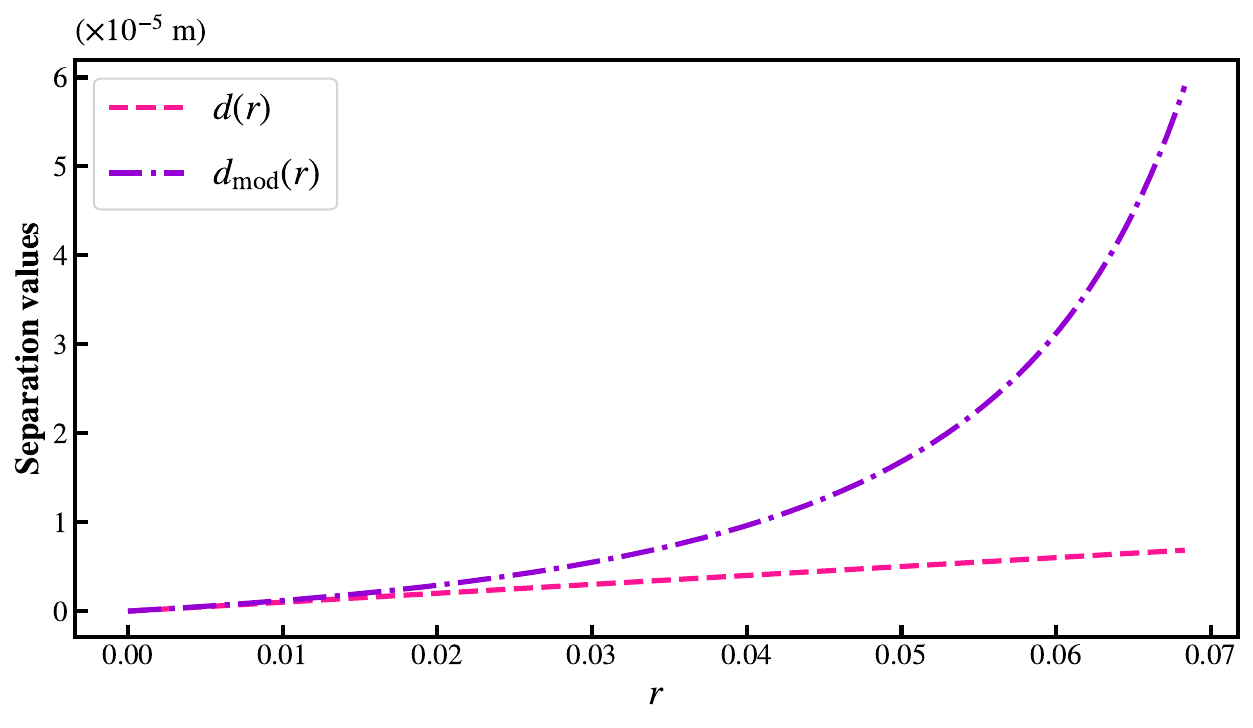}
\caption{Representative data for a comparison of superposition size ($d_{\rm mod}$, $d$) versus $r$ at $B_0 = 5~\mathrm{mT}$. $d_{\rm mod}$ 
incorporates the diamagnetic correction ($\eta \approx 11.04$), while $d$ neglects the diamagnetic interaction. Parameters: $m = 10^{10}$ amu, $w = 100\,\mu\mathrm{m}$, $\mu = \mu_{\rm NV}$.}
\label{fig:d&d'_maintext}
\vspace{-1.5mm}
\end{figure}

The time required for the lagging wavepacket to cross the confined magnetic field region of width w can give us an estimate of the time of evolution $t_f$ for a given $r$ and $\eta$ value. This in turn can help determine the maximum superposition size that can be generated while respecting the coherence time constraint $t_f < \tau$. 
\begin{equation}\label{eq:t_f_dia}
    t_f \approx w/v_+  = \frac{w}{v \sqrt{1-r\eta-r}}
\end{equation}
where $v = \sqrt{2K/M}$. Considering a NV-nanodiamond of mass $M=10^{10}$ and $B_0 = 5 ~mT$, $w = 100 ~\mu m$, this gives us $\eta = \mathcal{D/Z} = 11.04$ and at $r=0.077$, the total evolution time comes out as $t_f \approx 0.97 s < \tau = 1 s $, and the superposition size generated is $d_{mod} \approx 160 ~\mu m$. Increasing the mass and considering a NV-nanodiamond of mass $M=10^{12}$ along with $B_0 = 5 ~mT$, $w = 100 ~\mu m$, gives us $\eta = \mathcal{D/Z} = 1104.26$ and at $r= 8.3 \cross 10^{-4}$ the total evolution time comes out as $t_f \approx  0.95 s < \tau = 1 s $, and the superposition size generated is $d_{mod} \approx 3.5 ~\mu m$.

\subsection{Data} \label{app:data_NV}
\subsubsection{Evidencing MQC by varying incident velocity} \label{app:data_NV_vel}

\begin{table}[H]
\centering
\setlength{\tabcolsep}{9pt}
\renewcommand{\arraystretch}{1.0}
\begin{tabular}{cccccccc}
\toprule
$r$ & Temp (K) & $t_{f}$ (s) & $P'_{\chi}$ & $P_{\text{mixed}}$ & $P_{\chi}$ & $d$ (m) & $\mathcal{O}$ \\
\midrule
\num{1.00e-05} & \num{447.8092} & \num{9.47e-05} & 0.9684 & 0.5 & 0.9690 & \num{1.00e-09} & \num{0.9975} \\
\num{5.00e-05} & \num{89.5618}  & \num{2.12e-04} & 0.1521 & 0.5 & 0.1410 & \num{5.00e-09} & \num{0.9394} \\
\num{1.00e-04} & \num{44.7809}  & \num{3.00e-04} & 0.4599 & 0.5 & 0.4548 & \num{1.00e-08} & \num{0.7788} \\
\num{2.00e-04} & \num{22.3905}  & \num{4.24e-04} & 0.5083 & 0.5 & 0.5155 & \num{2.00e-08} & \num{0.3679} \\
\num{3.00e-04} & \num{14.9270}  & \num{5.19e-04} & 0.6270 & 0.5 & 0.8943 & \num{3.00e-08} & \num{0.1054} \\
\num{4.00e-04} & \num{11.1952}  & \num{5.99e-04} & 0.4337 & 0.5 & 0.0082 & \num{4.00e-08} & \num{0.0183} \\
\num{5.00e-04} & \num{8.9562}   & \num{6.70e-04} & 0.4941 & 0.5 & 0.3476 & \num{5.00e-08} & \num{0.0019} \\
\bottomrule
\end{tabular}
\caption{Data from analysis of sharp potential (see~\ref{app:analytical}): $m=10^{7}$~amu, $w$ = 100 $\mu\text{m}$, $B_0 = 5~\text{mT}$}
\label{tab:NV_anlyt_r}
\end{table}

\begin{table}[H]
\centering
\setlength{\tabcolsep}{9pt}
\renewcommand{\arraystretch}{1.0}
\begin{tabular}{cccccccc}
\toprule
$r$ & Temp (K) & $t_{f}$ (s) & $P'_{\chi}$ & $P_{\text{mixed}}$ & $P_{\chi}$ & $d$ (m) & $\mathcal{O}$ \\
\midrule
\num{1.00e-05} & \num{447.8092} & \num{1.61e-04} & 0.9684 & 0.5 & 0.9690 & \num{1.00e-09} & \num{0.9975} \\
\num{5.00e-05} & \num{89.5618}  & \num{3.60e-04} & 0.1520 & 0.5 & 0.1410 & \num{5.00e-09} & \num{0.9394} \\
\num{1.00e-04} & \num{44.7809}  & \num{5.09e-04} & 0.4599 & 0.5 & 0.4548 & \num{1.00e-08} & \num{0.7788} \\
\num{2.00e-04} & \num{22.3905}  & \num{7.19e-04} & 0.5085 & 0.5 & 0.5155 & \num{2.00e-08} & \num{0.3679} \\
\num{3.00e-04} & \num{14.9270}  & \num{8.81e-04} & 0.6272 & 0.5 & 0.8943 & \num{3.00e-08} & \num{0.1054} \\
\num{4.00e-04} & \num{11.1952}  & \num{1.02e-03} & 0.4337 & 0.5 & 0.0082 & \num{4.00e-08} & \num{0.0183} \\
\num{5.00e-04} & \num{8.9562}   & \num{1.14e-03} & 0.4940 & 0.5 & 0.3476 & \num{5.00e-08} & \num{0.0019} \\
\bottomrule
\end{tabular}
\caption{Data from approximate solution (see~\ref{app:approx}): $m=10^{7}$~amu, $s=0.1~(\mu\mathrm{m})^{-1}$ (smooth potential), $w$ = 100~$\mu\text{m}$, $B_0 = 5~\text{mT}$}
\label{tab:NV_num_r}
\end{table}

\begin{figure}[H]
    \centering
    \includegraphics[width=\textwidth]{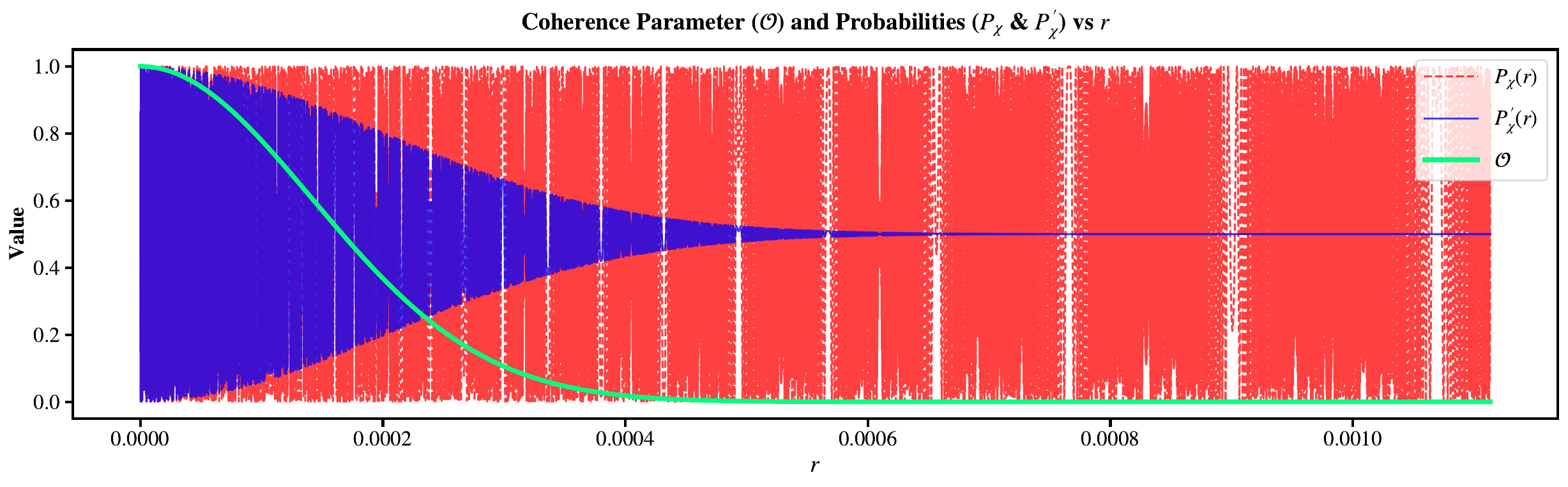}
    \caption{$\mathcal{O}$, $P_{\chi}$, $P^{\prime}_{\chi}$, $d$, and $d_{\rm mod}$ versus $r$. Parameters: $m = 10^{7}~\text{amu}$, $w = 100~\mu\mathrm{m}$, $\sigma = 10~\mathrm{nm}$, $B_{0} = 5~\text{mT}$.}
    \label{fig:plots_with_data_NV_1ms}
\end{figure}

\begin{figure}[H]
    \centering
    \includegraphics[width=0.55\textwidth]{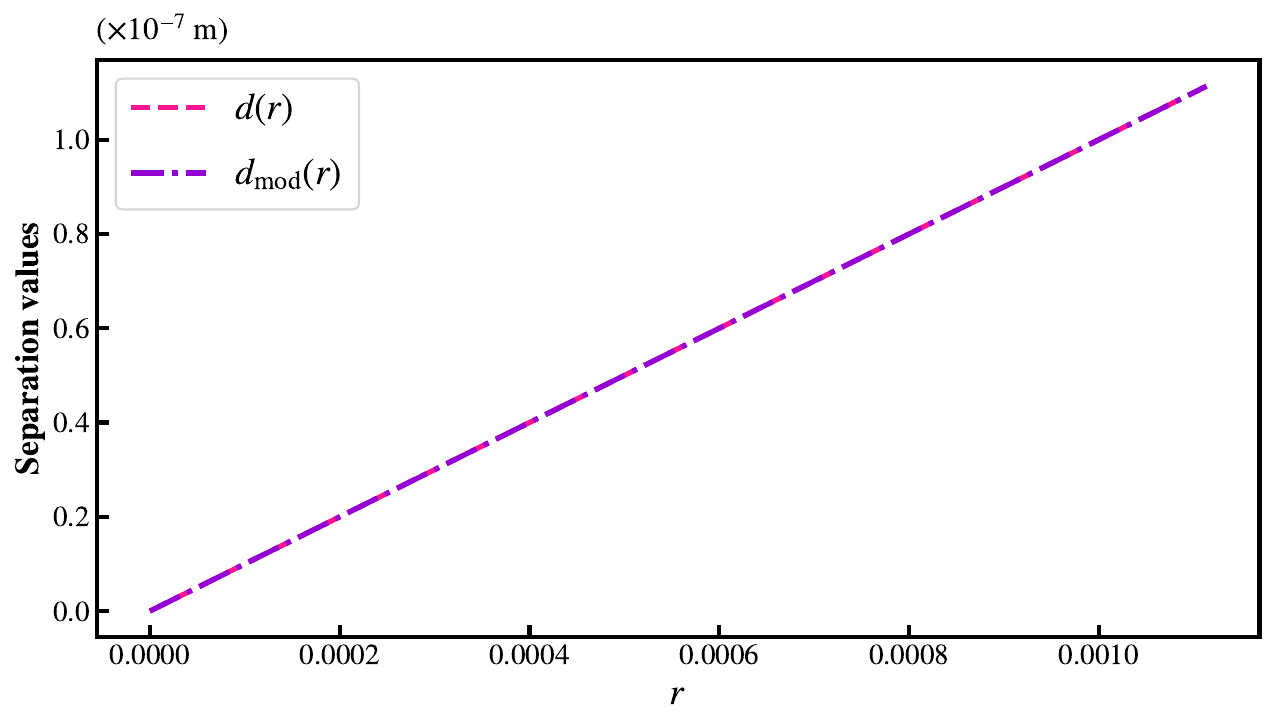}
    \caption{$\mathcal{O}$, $P_{\chi}$, $P^{\prime}_{\chi}$, $d$, and $d_{\rm mod}$ versus $r$. Parameters: $m = 10^{7}~\text{amu}$, $w = 100~\mu\mathrm{m}$, $\sigma = 10~\mathrm{nm}$, $B_{0} = 5~\text{mT}$.}
    \label{fig:d_plot_with_data_NV_1ms}
\end{figure}

\subsection{Discussion} \label{app:discussion_NV}

Firstly, note that since the sign of $\mu_{\rm NV}$ is opposite to that of $\mu_{\rm n}$, the dynamics of $\psi^{\pm}$ are interchanged in this case. Consequently, for a spin-embedded NV nanodiamond, the $\psi^{+}$ component emerges ahead of the $\psi^{-}$ component. The superposition size ($d$), coherence parameter ($\mathcal{O}$), spin probability ($P^{\prime}_{\chi}$), and other observable parameters remain unaffected. \vspace{2mm}

Secondly, we emphasize that the fall-off in $\mathcal{O}$ (and the oscillatory behavior of $P^{\prime}_{\chi}$) in the small-$r$ regime in this case is highly squished, corresponding to the small value of $\sigma = 10~\text{nm}$. This is consistent with the findings in~\ref{app:data1_sigma}. Thus, the definitions of ``small" and ``large" $r$ are also appropriately scaled in this case. As $r \to 1$, transmission starts to fall off oscillatorily and so does $P^{\prime}_{\chi}$. However, this large-$r$ regime may not be accessible for large masses owing to the coherence time constraint. Moreover, the effect of the falloff in transmission (and consequently $P^{\prime}_{\chi}$) is observed to be much suppressed for a smooth potential. Therefore, in an experimental scenario with a smooth potential, it would not be straightforward to observe this effect. \vspace{2mm}

Finally, we maintain that for these spin-embedded NV nanodiamond crystals, MQC is certified in the same manner as prescribed in the main paper, \textit{i.e.} by affirming the predicted trend for $\mathcal{O}$ vs $r$ and identifying spatial coherence by the criterion $0< \mathcal{O} < 1$. In addition, note that only a quantum coherent state can reproduce the predictions of $P^{\prime}_{\chi}$ in the \textit{complete} $r$ regime. Therefore, recording the value of the probability of spin outcome $\frac{1}{\sqrt{2}}\left( \ket{\uparrow}_z + \ket{\downarrow}_z \right)$, thus verifying the observed trend with our predictions and checking for $\lvert \Delta P_{\textrm{c}} \rvert \neq 0$ \textit{and} $P^{\prime}_{\chi} \neq 0.5$ is another route to certify the spatial coherence in the emergent state.

\subsection{Near-future possibility with advances in coherence time} \label{app:1s_coherence_NV}

The duration of the experiment is most notably constrained by the NV electronic spin coherence time. The longest reported coherence times in nanodiamonds reach $786~\mu\text{s}\approx 1~\text{ms}$ at room temperature by applying dynamical decoupling~\cite{NV_coherence_time18,NV_coherence_time22,NV_coherence_time23}. These limits are likely set by residual nitrogen concentrations ($\sim$100 ppb) and unpaired surface spins in $^{12}\mathrm{C}$ nanodiamonds. Further reductions in nitrogen density and increasing the diamond size to the micron scale (e.g.~via nanopillar fabrication from electronic-grade $^{12}\mathrm{C}$ diamond) are expected to enhance coherence. As a matter of fact, NV spin coherence times exceeding $1~\text{s}$ have been demonstrated at cryogenic temperatures~\cite{1s_coherence_Abobeih2018,1s_coherence_PhysRevX.9.2019}, suggesting similar gains for high-purity nano- and microdiamonds. An additional advantage of high-purity nanodiamonds is their increased optical transparency, which reduces absorption of the green excitation and infrared tracking light, thereby mitigating heating, an especially important consideration under cryogenic operation. In this context, continued technological advances are expected to significantly relax the constraints imposed by spin coherence times. With this consideration, taking an improved spin-coherence time upto $\tau = 1~\text{s}$, we present corresponding results from our coupled treatment, demonstrating a substantial enhancement in the maximum achievable mass of the superposed macroscopic system.

\subsubsection{Data: Scaling up mass} \label{app:1s_coh_NV_m}

\begin{table}[h]
\centering
\setlength{\tabcolsep}{10pt}
\renewcommand{\arraystretch}{1.0}
\begin{tabular}{c|cccccc}
\toprule
$m$ (amu) & $t_{f}$ (s) & $P'_{\chi}$ & $P_{\text{mixed}}$ & $P_{\chi}$ & $d$ (m) & $\mathcal{O}$ \\
\midrule
$10^6$ & \num{5.99e-3} & 0.5478 & 0.5 & 0.8529 & \num{4e-8} & \num{0.0183} \\
$10^7$ & \num{1.89e-2} & 0.5162 & 0.5 & 0.6204 & \num{4e-8} & \num{0.0183} \\
$10^8$ & \num{5.99e-2} & 0.4986 & 0.5 & 0.4910 & \num{4e-8} & \num{0.0183} \\
$10^9$ & \num{1.89e-1} & 0.5511 & 0.5 & 0.8797 & \num{4e-8} & \num{0.0183} \\
$10^{10}$ & \num{5.99e-1} & 0.4337 & 0.5 & 0.0082 & \num{4e-8} & \num{0.0183} \\
\bottomrule
\end{tabular}
\caption{Representative data from analysis of sharp potential (see~\ref{app:analytical}): $r=\num{4e-4}$, $w$ = 100 $\mu\text{m}$, $B_0 = 5~\mu\text{T}$, Temp. = $0.0112~\text{K}$}
\label{tab:NV_anlyt_m_r=4e-4}
\end{table}

\begin{table}[h]
\centering
\setlength{\tabcolsep}{10pt}
\renewcommand{\arraystretch}{1.0}
\begin{tabular}{c|cccccc}
\toprule
$m$ (amu) & $t_{f}$ (s) & $P'_{\chi}$ & $P_{\text{mixed}}$ & $P_{\chi}$ & $d$ (m) & $\mathcal{O}$ \\
\midrule
$10^6$ & \num{9e-3} & 0.5478 & 0.5 & 0.8529 & \num{4e-8} & \num{0.0183} \\
$10^7$ & \num{2.9e-2} & 0.5163 & 0.5 & 0.6204 & \num{4e-8} & \num{0.0183} \\
$10^8$ & \num{9.1e-2} & 0.4987 & 0.5 & 0.4910 & \num{4e-8} & \num{0.0183} \\
$10^9$ & \num{2.9e-1} & 0.5511 & 0.5 & 0.8797 & \num{4e-8} & \num{0.0183} \\
$10^{10}$ & \num{9.7e-1} & 0.4337 & 0.5 & 0.0082 & \num{4e-8} & \num{0.0183} \\
\bottomrule
\end{tabular}
\caption{Representative data from approximate solution (see~\ref{app:approx}) $r=\num{4e-4}$, $s=0.1~(\mu\mathrm{m})^{-1}$ (smooth potential), $w$ = 100 $\mu\text{m}$, $B_0 = 5~\mu\text{T}$, Temp. = $0.0112~\text{K}$}
\label{tab:NV_num_m_r=4e-4}
\end{table}

\newpage
\subsubsection{Data: Evidencing MQC for $m=10^{10}~\text{amu}$} \label{app:1s_coh_NV_vel}

\begin{table}[h]
\centering
\setlength{\tabcolsep}{9pt}
\renewcommand{\arraystretch}{1.0}
\begin{tabular}{cccccccc}
\toprule
$r$ & Temp (K) & $t_{f}$ (s) & $P'_{\chi}$ & $P_{\text{mixed}}$ & $P_{\chi}$ & $d$ (m) & $\mathcal{O}$ \\
\midrule
\num{1.00e-06} & \num{4.4780} & 0.03 & 0.1497 & 0.5 & 0.1496 & \num{1.00e-10} & \num{1} \\
\num{5.00e-06} & \num{0.8956} & 0.07 & 0.0090 & 0.5 & 0.0088 & \num{5.00e-10} & \num{0.9994} \\
\num{1.00e-05} & \num{0.4478} & 0.09 & 0.9684 & 0.5 & 0.9690 & \num{1.00e-09} & \num{0.9975} \\
\num{5.00e-05} & \num{0.0896} & 0.21 & 0.1521 & 0.5 & 0.1410 & \num{5.00e-09} & \num{0.9394} \\
\num{1.00e-04} & \num{0.0448} & 0.30 & 0.4599 & 0.5 & 0.4548 & \num{1.00e-08} & \num{0.7788} \\
\num{2.00e-04} & \num{0.0224} & 0.42 & 0.5083 & 0.5 & 0.5156 & \num{2.00e-08} & \num{0.3679} \\
\num{3.00e-04} & \num{0.0149} & 0.52 & 0.6270 & 0.5 & 0.8943 & \num{3.00e-08} & \num{0.1054} \\
\num{4.00e-04} & \num{0.0112} & 0.60 & 0.4337 & 0.5 & 0.0082 & \num{4.00e-08} & \num{0.0183} \\
\num{4.50e-04} & \num{0.0099} & 0.64 & 0.5276 & 0.5 & 0.8366 & \num{4.50e-08} & \num{0.0063} \\
\num{5.00e-04} & \num{0.0090} & 0.67 & 0.4940 & 0.5 & 0.3476 & \num{5.00e-08} & \num{0.0019} \\
\num{1.00e-03} & \num{0.0045} & 0.96 & 0.5 & 0.5 & 0.0396 & \num{1.00e-07} & \num{0} \\
\bottomrule
\end{tabular}
\caption{Representative data from analysis of sharp potential (see~\ref{app:analytical}): $m=10^{10}$ amu, $w$ = 100 $\mu\text{m}$, $B_0 = 5~\mu\text{T}$}
\label{tab:NV_anlyt_r_10^10amu}
\end{table}

\begin{table}[h]
\centering
\setlength{\tabcolsep}{9pt}
\renewcommand{\arraystretch}{1.0}
\begin{tabular}{cccccccc}
\toprule
$r$ & Temp (K) & $t_{f}$ (s) & $P'_{\chi}$ & $P_{\text{mixed}}$ & $P_{\chi}$ & $d$ (m) & $\mathcal{O}$ \\
\midrule
\num{1.00e-06} & \num{4.4780} & 0.05 & 0.1497 & 0.5 & 0.1496 & \num{1.00e-10} & \num{1} \\
\num{5.00e-06} & \num{0.8956} & 0.11 & 0.0090 & 0.5 & 0.0088 & \num{5.00e-10} & \num{0.9994} \\
\num{1.00e-05} & \num{0.4478} & 0.15 & 0.9684 & 0.5 & 0.9690 & \num{1.00e-09} & \num{0.9975} \\
\num{5.00e-05} & \num{0.0896} & 0.32 & 0.1520 & 0.5 & 0.1410 & \num{5.00e-09} & \num{0.9394} \\
\num{1.00e-04} & \num{0.0448} & 0.45 & 0.4600 & 0.5 & 0.4548 & \num{1.00e-08} & \num{0.7788} \\
\num{2.00e-04} & \num{0.0224} & 0.64 & 0.5085 & 0.5 & 0.5156 & \num{2.00e-08} & \num{0.3679} \\
\num{3.00e-04} & \num{0.0149} & 0.78 & 0.6272 & 0.5 & 0.8943 & \num{3.00e-08} & \num{0.1054} \\
\num{4.00e-04} & \num{0.0112} & 0.90 & 0.4337 & 0.5 & 0.0082 & \num{4.00e-08} & \num{0.0183} \\
\num{4.50e-04} & \num{0.0099} & 0.96 & 0.5275 & 0.5 & 0.8366 & \num{4.50e-08} & \num{0.0063} \\
\num{5.00e-04} & \num{0.0090} & 1.02 & 0.4940 & 0.5 & 0.3476 & \num{5.00e-08} & \num{0.0019} \\
\num{1.00e-03} & \num{0.0045} & 1.45 & 0.5 & 0.5 & 0.0396 & \num{1.00e-07} & \num{0} \\
\bottomrule
\end{tabular}
\caption{Representative data from approximate solution (see~\ref{app:approx}): $m=10^{10}$ amu, $s=0.1~(\mu\mathrm{m})^{-1}$ (smooth potential), $w=100~\mu$m, $B_0=5~\mu\text{T}$.}
\label{tab:NV_num_r_smooth_10^10amu}
\end{table}

\begin{figure}[H]
    \centering
    \includegraphics[width=\textwidth]{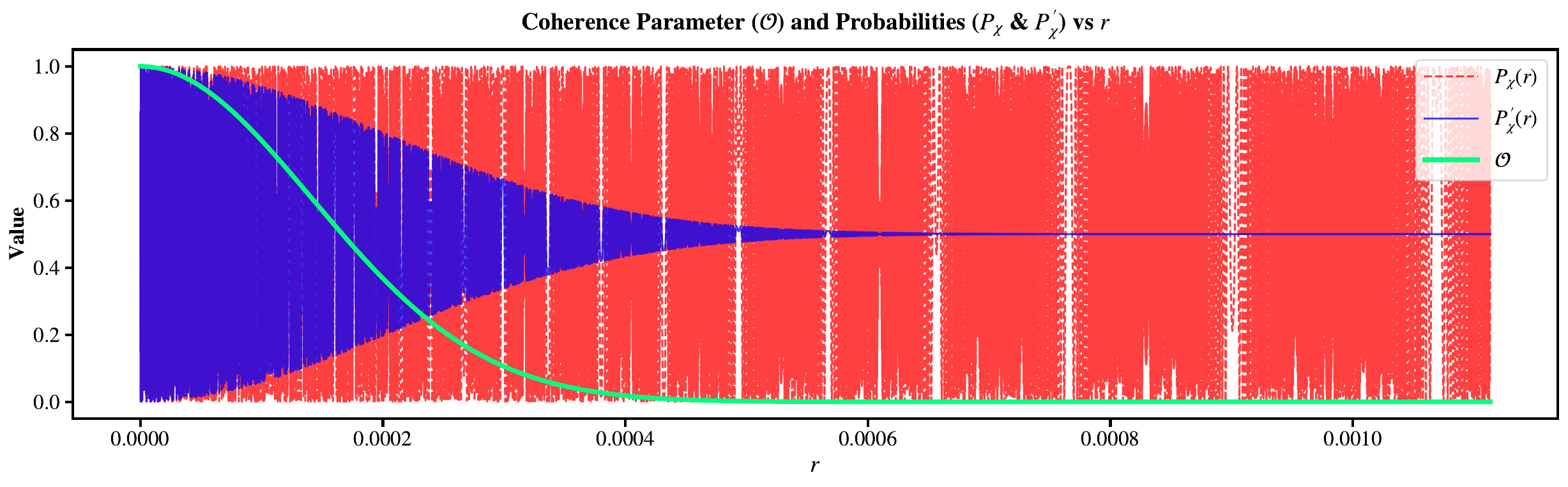}
    \caption{$\mathcal{O}$, $P_{\chi}$, and $P^{\prime}_{\chi}$ versus $r$. Parameters: $m = 10^{10}~\text{amu}$, $w = 100~\mu\mathrm{m}$, $\sigma = 10~\mathrm{nm}$, $B_{0} = 5~\mu\text{T}$.}
    \label{fig:plots_with_data_NV_1s_p1}
\end{figure}
\begin{figure}[H]
    \centering
    \includegraphics[width=0.55\textwidth]{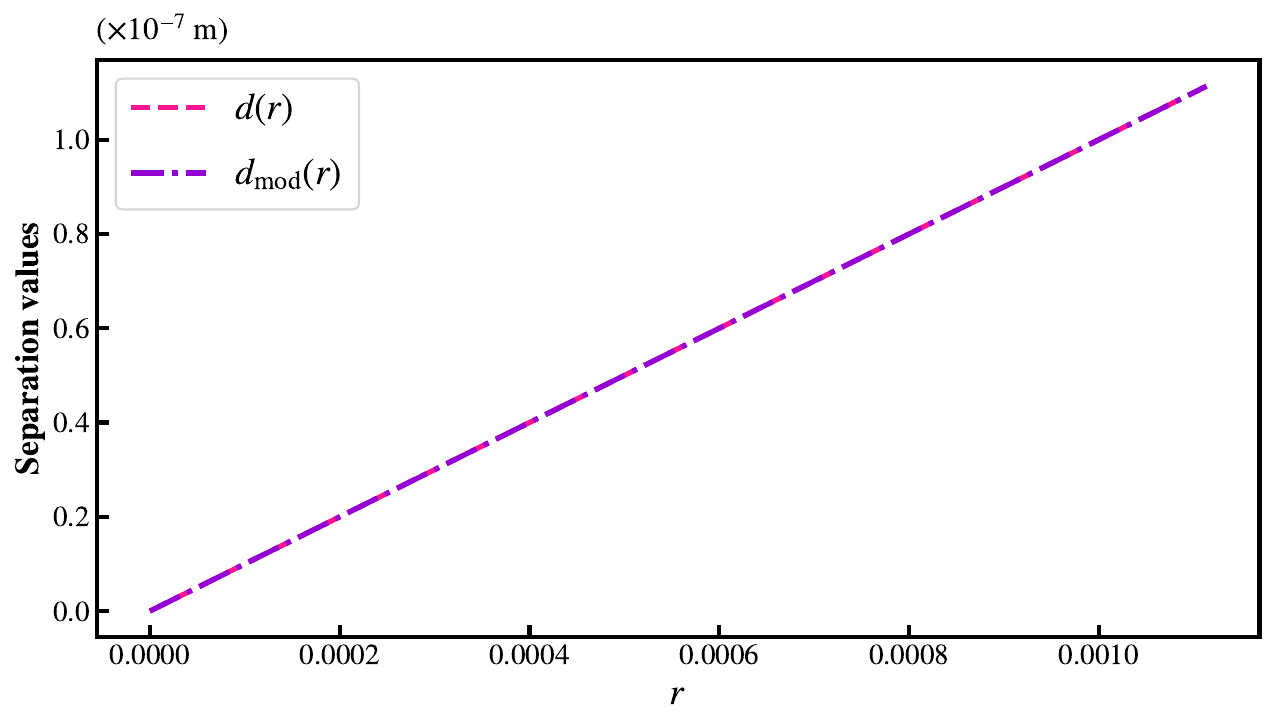}
    \caption{$d$ and $d_{\rm mod}$ versus $r$. Parameters: $m = 10^{10}~\text{amu}$, $w = 100~\mu\mathrm{m}$, $\sigma = 10~\mathrm{nm}$, $B_{0} = 5~\mu\text{T}$.}
    \label{fig:plots_with_data_NV_1s_p2}
\end{figure}

\newpage
\section{Potential sensing application} \label{app:sensing}

To begin with, we note that conventional NMR-based magnetometers harness Larmor precession of NV-center spin ensembles to estimate field strengths, achieving $\sim$picoTesla sensitivities and sub-100 nanometer-scale spatial resolution~\cite{qsenserev,B_sensing_90,B_sensing_11,barry_sensing2020}. We point out how our framework finds direct relevance in this context. \vspace{1mm}

Firstly, existing magnetometers operating in the kinetic regime~\cite{B_sensing_08} and relying on textbook semiclassical Larmor precession would require to correct their estimates to account for the deviations predicted by our full quantum (kinetic) treatment of the situation. Secondly, our analysis suggests an alternative means of characterizing both the strength and spatial extent of an external magnetic field. Since, in the kinetic case, the emergent state naturally assumes a coherent spatial superposition, it offers the intriguing possibility of using the spatial separation ($d$) between the $\psi^{\pm}$ components to quantify the magnetic field strength and/or the spatial extent of the field. Eq.~\ref{eq:c16} can be expanded with respect to $r$ and a ratio of two different values of $d$ (essentially for two different kinetic energies) can be taken to remove the dependence on $w$.
\begin{equation} \label{eq:series_d}
    d = w \left( r + \frac{5}{8}\, r^{3} + \frac{63}{128}\, r^{5} + \frac{429}{1024}\, r^{7} + \cdots \right)\, ,
\qquad |r| < 1
\end{equation}
Experimentally, this translates to measuring the value of separation parameter ($d_1$, $d_2$) for two different kinetic energy values ($K_1$, $K_2$) and from this the field strength can be determined. Notice that only odd powers of $r = |\mu B_{0}|/K$ survive the expansion. By truncating after the cubic term, the solution for $B_{0}$ simply reads,
\begin{equation} \label{eq:B_sol_r^3}
  B_{0} = \sqrt{\frac{d_1 K_1 - d_2 K_2}{\tfrac{5}{8} \mu^2 \left(\tfrac{d_2}{K_2}-\tfrac{d_1}{K_1}\right)}} \, .
\end{equation}
One can obtain an analytical solution for $B_{0}$ (and the corresponding $w$) by expanding upto maximum $r^{5}$ order terms (higher order than that is not analytically tractable in accordance with the Abel-Ruffini Theorem). Within an order-5 approximation, we essentially solve for $B_{0}$ from the following equation,
\begin{equation} \label{eq:d_ratio_r^5}
    \frac{d_1}{d_2} = \left(\frac{K_2}{K_1}\right)^5 \, \frac{128 K_1^4 + 80 K_1^2 \mu^2 B_{0}^{2} + 63 \mu^4 B_{0}^{4}} {128 K_2^4 + 80 K_2^2 \mu^2 B_{0}^{2} + 63 \mu^4 B_{0}^{4}} \, .
\end{equation}
The solution is,
\begin{equation} \label{eq:B_sol_r^5}
B_{0} = \sqrt{\frac{\mathcal{N}}{63 \mu^{2} \left( K_1^{5} d_1 - K_2^{5} d_2 \right)}} \, ,
\end{equation}
such that,
\begin{equation} \label{eq:factors_B_sol_r^5}
\begin{aligned}
    &\mathcal{N} = -40 K_1^{5} K_2^{2} d_1 + 40 K_1^{2} K_2^{5} d_2 + 8 K_1^{2} K_2^{2} \sqrt{\mathcal{D}} \,, \quad \text{where} \\
    &\mathcal{D} = -101 K_1^{6} d_1^{2} + 126 K_1^{5} K_2 d_1 d_2 - 50 K_1^{3} K_2^{3} d_1 d_2 + 126 K_1 K_2^{5} d_1 d_2 - 101 K_2^{6} d_2^{2} \,.
\end{aligned}
\end{equation}
Then, the spatial extent of the field can be recovered from,
\begin{equation} \label{eq:w_sol}
    w = \frac{d}{\left(\sqrt{\frac{K}{K - |\mu B_{0}|}} - \sqrt{\frac{K}{K + |\mu B_{0}|}}\right)} \, .
\end{equation}
Thus, the emergent spatial modes may be viewed as effective “spatial qubits”~\cite{massive_qubits_Bin23}, where position-resolved measurements yielding the value of separation $d$ directly reveals the underlying field characteristics. Since $d \propto w$, the ability to discern smaller separations corresponds to higher spatial resolution. This approach could enable the probing of uniform magnetic field regions over very small length scales. \vspace{2mm}

\noindent We present a sample calculation based on these findings, to verify how accurate these are to the true values of the field strength $B_{0}$ and the spatial extent of the field $w$. We take a realistic set of parameters for an NV center spin-embedded nanodiamond: $\mu_{\rm NV} = 1.8548020157\times 10^{-23}~\text{J\,.\,T}^{-1}$, $\sigma = 10~\text{nm}$, and $m = 10^{6}~\text{amu}$. The magnetic field is initialized as $B_{0} = 10^{-6}~\text{T}$, and $w = 1~\mu\text{m}$. Using these parameters, we compute the value of the separation $d$ at two different values of the kinetic energy $K$, corresponding to two values of $r$ for a fixed mass. For example, using $r_{1}=0.5$ and $r_{2}=0.7$, we recover $B_{0} = \num{0.98e-06}~\text{T}$ and $w = 1.03~\mu\text{m}$.

\end{document}